\documentclass[twocolumn,aps,prl,showpacs,amsmath,superscriptaddress,longbibliography,notitlepage]{revtex4-2}
\usepackage{amssymb}
\usepackage{mathrsfs}
\usepackage{graphicx}
\usepackage{float}
\usepackage[caption=false]{subfig}
\usepackage[pdftex,colorlinks=red]{hyperref}

\usepackage{amsmath}
\usepackage{epstopdf}
\usepackage[normalem]{ulem}
\usepackage{verbatim}
\usepackage{xcolor}
\usepackage{bm}

\def\blue{\textcolor{black}}

\def\LLH{\textcolor{black}}

\begin{document}

\title{Topological and fractal defect states in non-Hermitian lattices}
\author{Gan Liang}
\affiliation{Guangdong Provincial Key Laboratory of Quantum Metrology and Sensing $\&$ School of Physics and Astronomy, Sun Yat-Sen University (Zhuhai Campus), Zhuhai 519082, China}
\author{Linhu Li}\email{lilh56@mail.sysu.edu.cn}
\affiliation{Guangdong Provincial Key Laboratory of Quantum Metrology and Sensing $\&$ School of Physics and Astronomy, Sun Yat-Sen University (Zhuhai Campus), Zhuhai 519082, China}
\affiliation{Quantum Science Center of Guangdong-Hong Kong-Macao Greater Bay Area (Guangdong), Shenzhen, China}

\begin{abstract}
Higher dimensions provide fertile ground for diverse topological phases and their associated localization phenomena, thanks to the rich geometric features of boundaries and defects. In this paper, we investigate non-Hermitian lattices with defects and establish a correspondence between spectral winding topology, fractal structures, and defect-localized states in arbitrary dimensions. Through analytical derivation and numerical simulations, we demonstrate that defect states emerge only when the spectral winding number exceeds a threshold determined by the defect size, which is linked to their fractal characteristics. By utilizing the Green's function, we identify amplified responses at defects under external driving fields, strengthening the physical correspondence between these topological and fractal features. Our findings offer a universal framework for understanding defect-localized states in higher-dimensional non-Hermitian systems.
\end{abstract}
\maketitle
\emph{Introduction.}---
Topological phases of matter have become a focal point in contemporary physics, largely due to the robust localization of states that arise when translational symmetry is broken by boundaries or defects~\cite{hasan2010colloquiuma,qi2011topological,RevModPhys.88.035005,teo2010topological}. In recent years, attention has increasingly turned to non-Hermitian Hamiltonians describing non-conservative systems~\cite{bender1998real,rotter2009non,bender2007making,nonH_review1,nonH_review2,nonH_review3,nonH_review4,nonH_review5,nonH_review6}. These systems host a unique class of point-gap topological phases characterized by a spectral winding number~\cite{kawabata2019symmetry}, leading to anomalous localization phenomena such as the non-Hermitian skin effect (NHSE)~\cite{PhysRevLett.121.136802,alvarez2018nonH,yao2018edge,borgnia2020nonH,okuma2020topological,zhang2020correspondence} and the scale-free localization~\cite{li2020critical,li2021impurity,guo2021exact}. 
\blue{Alternative real-space methods have also been applied to investigate NHSE, in order to diagnose its topological nature even when translational symmetry is broken by boundaries, disorders, or defects~\cite{claes2021skin,schomerus2023renormalization,chadha2024real,banerjee2024non}.}
In higher dimensions, non-Hermitian systems present a much richer variety of state localization patterns driven by the geometric features of boundaries and defects~\cite{lee2019hybrid,li2020topological,li2022gainlossinduced,zhu2022hybrid, sun2021geometric,schindler2021dislocation,panigrahi2022non,bhargava2021non,manna2023inner,ou2023nonH}.

A central and often overlooked question in this field concerns the full physical consequences of spectral winding topology, a $Z$-type topology with integer invariants. 
While a nonzero spectral winding number explains the NHSE, 
the precise correspondence to different integers of the winding number, particularly for higher-dimensional systems, remains unclear. 
Limited studies have explored these issues only in one-dimensional systems with semi-infinite~\cite{okuma2020topological,pan2021point} or continuously varying boundary conditions~\cite{li_quantized_2021,PhysRevB.105.L241402}.
In this work, we uncover a class of defect-localized states in non-Hermitian systems that are characterized by the exact values of spectral winding numbers.

Through analytical methods and numerical verifications, we demonstrate that 
\blue{in a $m$-dimensional ($m$D) lattice with $(m-1)$D defects,}
defect states emerge when a spectral winding number exceeds a critical threshold determined by the size of the defects,
\blue{representing a precise physical correspondence of the spectral winding topology in non-Hermitian systems}.
This correspondence is independent of dimensionality or crystal symmetry, establishing a universal framework for characterizing non-Hermitian topological defect states \blue{in arbitrary dimensions}.
By further taking into account fractal structures of defects, we establish a correspondence between their fractal dimension and winding topology through the emergence of these skin defect states.
Furthermore, we show that these states manifest as signal amplification at defects in steady-state response to external driving fields, providing a clear physical signature for detecting the topological and fractal features in both classical and quantum platforms.

\emph{\blue{$(m-1)$D skin defect states in a $m$D lattice.}}---
\begin{figure}
    \includegraphics[width=0.48\textwidth]{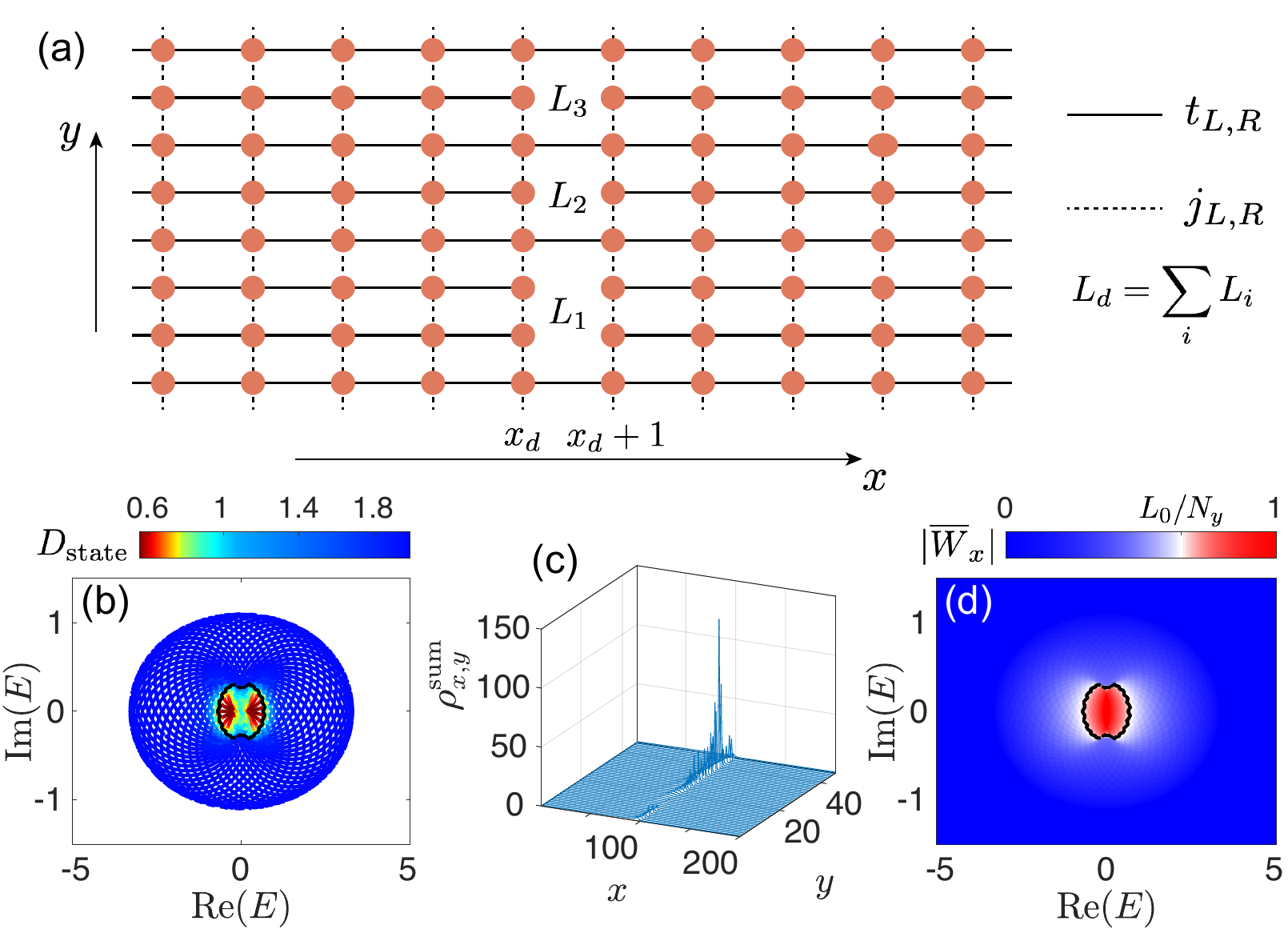}  
    \caption{
    (a) A sketch of the 2D lattice described by Eq.~\eqref{eq:Hamiltonian2D}, with defects generated by removing the hopping between certain lattice sites at $x_d$ and $x_{d+1}$.
    $L_i$ with $i=1,2,3,...$ represents the length of each segment of the defects, and $L_d$ denotes their total length.
    (b) Eigenenergies of the defective model, marked by different colors according to $D_{\rm state}$, the fractal dimension of eigenstates.
    (c) Summed distribution of all eigenstates. (c1)A strong localization at the defects is observed.
    (d) Absolute values of the average spectral winding number. 
    Black dots separate the regions with $|\overline{W}_x|$ larger and smaller than $L_0/N_y$ (with $L_0=N_y-L_d$), which consist with the boundary between localized and extended states in (b).
    In (b) to (d), parameters are
    $t_R=1.2$, $t_L=0.5$, $j_R=0.6$, $j_L=1$. $N_x=200$, $N_y=47$, $x_d=100$,
    with $16$ defective sites along $y$ direction, located at $y=3,5,7,9,15,23,25,27,29,32,35,37,39,40,44,45$.
    }
    \label{fig:model}
\end{figure}
We start with 
\blue{a minimal $m$D non-Hermitian lattice with nearest-neighbor hoppings and $(m-1)$D defects.
For clarity, we label the dimensions as $x,\mathbf{y}=(y_1,y_2,...,y_{m-1})$, with the defects embedded in a $(m-1)$D surface of $\mathbf{y}$, cutting off the hopping between sites at $x_d$ and $x_d+1$.
Its Hamiltonian is given by (without defects)
\begin{eqnarray}
    \label{eq:Hamiltonian2D}
    H&=&\sum_{x}^{N_x}\sum_{\mathbf{y}}^{\mathbf{N_y}}(t_R\hat{c}^{\dagger}_{x+1,\mathbf{y}}\hat{c}_{x,\mathbf{y}}+t_Lc^{\dagger}_{x,\mathbf{y}}\hat{c}_{x+1,\mathbf{y}})\\
   &&+\sum_i^{m-1} \sum_{x}^{N_x}\sum_{\mathbf{y}}^{\mathbf{N_y}}(j_{R,y_i} \hat{c}^{\dagger}_{x,\mathbf{y}+\bm{\delta}_i}\hat{c}_{x,\mathbf{y}}+j_{L,y_i} \hat{c}^{\dagger}_{x,\mathbf{y}}\hat{c}_{x,\mathbf{y}+\bm{\delta}_i}),\nonumber
\end{eqnarray}
with $\hat{c}_{x,\mathbf{y}}$ the annihilation operator of a particle at lattice site $(x,\mathbf{y})$,
$\bm{\delta}_i=(0_1,0_2,...1_i,...,0_{m-1})$,
$N_x$ and $\mathbf{N_y}=(N_{y_1},N_{y_2},...,N_{y_{m-1}})$ the lattice sizes along different directions,
and $t_{R/L}$ ($j_{R/L,y_i}$) the asymmetric hopping amplitudes along $x$ ($y_i$) direction.
Here we have chosen periodic boundary conditions (PBCs) along all directions; however, as can be seen later, the boundary conditions along $\mathbf{y}$ are irrelevant to our formalism.}

\blue{As a numerical demonstration, we consider the simplest 2D case with $\mathbf{y}=y$, $j_{R/L,y_1}\equiv j_{R/L}$, and line defects with a total length $L_d$ along $y$ direction, as sketched in Fig. \ref{fig:model}(a).}
The defects break the translational symmetry along $x$ and are expected to host localized states manifesting the NHSE.
However, only a portion of eigenstates are found to be localized at defects,
in contrast to conventional NHSE in this model where all eigenstates become localized. 
In Fig. \ref{fig:model}(b), we demonstrate the eigenenergies  marked by colors according to \blue{an effective} dimension of their corresponding eigenstates,
defined as 
\begin{eqnarray}
    \label{eq:IPR}
 D_{\rm state}=-\log \left[\sum_{x,y} |\psi_{x,y}|^4\right]/\log\sqrt{N_xN_y},
\end{eqnarray}
with $\psi_{x,y}$ the wave amplitude at site $(x,y)$ of a \blue{normalized} eigenstate $|\Psi\rangle$.
\blue{We note that $D_{\rm state}$ coincides with the fractal dimension of the state in the thermodynamic limit ($N_x,N_y\rightarrow \infty$), where $D_{\rm state}=2$ for extended bulk states, and $D_{\rm state}=1$ for states localized at the 1D defects.
In a finite-size system, defect states in our model shall generally have $D_{\rm state}<1$ since $L_d<N_y$.}
The total distribution summed over all eigenstates, $\rho^{\rm sum}_{x,y}=\sum |\psi_{x,y}|^2$, is shown in Fig. \ref{fig:model}(c), where a strong localization is seen along the defective lattices.
Note that the skin defect states are seen to distribute non-uniformly along $y$ direction, 
due to the non-reciprocal hopping amplitudes ($j_R\neq j_L$) and the lack of translation symmetry of defects along $y$ direction.



\emph{Formal solution of the lattice.}---
To unravel the origin and anomalous properties of the skin defect states,
we first solve the formal solutions for an equivalent 1D Hamiltonian where
$\mathbf{y}$ directions are taken as an internal degree of freedom,
\begin{eqnarray}
    \label{eq:Hamiltonian}
    H\rightarrow H_{\rm 1D}&=&\sum_{x=1}^{N_{\rm x}-1}(\hat{C}^{\dagger}_{x+1}J_R\hat{C}_{x}+\hat{C}^{\dagger}_{x}J_L\hat{C}_{x+1})
    \\
    &+&(\hat{C}^{\dagger}_{1}J_RB\hat{C}_{N_{\rm x}}+\hat{C}^{\dagger}_{N_{x}}J_LB\hat{C}_{1})
    +\sum_{x=1}^{N_{\rm x}}(\hat{C}^{\dagger}_xJ\hat{C}_x),
    \nonumber
\end{eqnarray}
\blue{with the column vector of annihilation operators
$\hat{C}_x=\left[\cdots~c_{x,\mathbf{y}}~\cdots\right]^{\rm T}$.
The elements of hopping matrices $J_{R/L}$ and $J$ are given by
$(J_{R/L})_{\mathbf{y}_1,\mathbf{y}_2}=t_{R/L}\delta_{\mathbf{y}_1,\mathbf{y}_2}$ and
$(J)_{\mathbf{y}_1,\mathbf{y}_2}=\sum_{i}\left(j_{L,y_i}\delta_{\mathbf{y}_1+\bm{\delta}_i,\mathbf{y}_2}+j_{R,y_i}\delta_{\mathbf{y}_1,\mathbf{y}_2+\bm{\delta}_i}\right)$, respectively.
$B$ is a diagonal matrix describing the defects,
with $(B)_{n,n}\equiv b_{n}$ take $0$ ($1$) if the lattice site with $\mathbf{y}\rightarrow n$ is at (away from) the defects.}
Without loss of generality, we can place the line defects as the boundary of the system, by setting $x_d+1\equiv 1$ and $x_d\equiv N_x$.

We now provide the key steps of our derivation, with more details in Supplemental Materials~\cite{SupMat}.
We first diagonalize the coupling matrix $J$ as $Q^{-1}JQ=\Lambda$ with $\Lambda$ a diagonal matrix.
\blue{Since $J_{R}$ and $J_L$ commute with $J$,}
the eigenequation $H|\Psi\rangle=E|\Psi\rangle$, 
with $|\Psi\rangle=\sum_{x,\mathbf{y}}\psi_{x,\mathbf{y}}\hat{c}^\dagger_{x,\mathbf{y}}|0\rangle$ and $|0\rangle$ the vaccum state,
can be expressed as
\begin{eqnarray}
    \label{eq:transformed_bulk}
    &&E\Phi_x=\Lambda\Phi_x+t_R\Phi_{x-1}+t_L\Phi_{x+1}\\
    \label{eq:transformed_boundary}
    &&Q\Phi_0=BQ\Phi_{N_x},~~
    Q\Phi_{N_{x}+1}=BQ\Phi_{1},
\end{eqnarray}
in the bulk and at the defect, respectively,
with
\begin{eqnarray}
    \label{eq:phi}
    Q^{-1}\Psi_x\equiv \Phi_x=
    \begin{bmatrix}
        \phi _{x,1}&
        \phi _{x,2}&
        \cdots&   
        \phi _{x,y'}&       
        \cdots&       
        \phi _{x,N_y}
       \end{bmatrix}^{\rm T},
\end{eqnarray}
and each $\phi_{x,y'}$ a linear combination of wave amplitudes at different $\mathbf{y}$.
\blue{Note that for simplifying the notations, we have mapped the $(m-1)D$ of $\mathbf{y}$ space to a 1D space with $y'=1,2,...,N_y$ and $N_y\equiv\prod_{i=1}^{m-1}N_{y_i}$. }

As $\Lambda$ is diagonalized, 
the bulk equations of Eq.~\eqref{eq:transformed_bulk} are decoupled for different $y'$, while the defect equations of Eq.~\eqref{eq:transformed_boundary} are not.
Thus we can treat each component $\phi_{x,y‘}$ separately in the bulk, and obtain a formal solution as for the 1D Hatano-Nelson model~\cite{hatano_nelson1,hatano_nelson2},
\begin{eqnarray}
    \label{eq:solution_form}
    \phi_{x,y'}=\gamma_{+,y'}\beta_{+,y'}^x+\gamma_{-,y'}\beta_{-,y'}^x,~
    \beta_{\pm,y'}=Rr_{y'}^{\pm1}e^{\mp i\alpha_{y'}},\nonumber\\
\end{eqnarray}
with $R=\sqrt{{t_R}/{t_L}}$,  $r_{y'}$ and $\alpha_{y'}$ being real, and $\gamma_{\pm}$ the coefficients determined by boundary conditions~\cite{guo2021exact,SupMat}.
\blue{Note that this formal solution, and later discussion of topological origin of skin defect states, rely on the commutation relation $[J_{R/L} ,J]=0$. 
Otherwise,
more sophisticated analyses are required to obtain the same formalism of topological correspondence for skin defect states, as discussed in Sec. V in the Supplemental Materials~\cite{SupMat}.}

\emph{Topological origin of the skin defect states.}---
Without loss of generality, we set $t_R>t_L$ (so that $R>1$) and $r_{y'}>1$ in the rest of this paper,
where skin defect states shall localize toward $x=N_x$.
Following Eq.\eqref{eq:phi} and Eq.\eqref{eq:solution_form}, these states must be composed by a series of $\gamma_{\pm,y'}$ with $|\beta_{\pm,y'}|>1$.
In our convention, $|\beta_{+,y'}|>1$ is always satisfied,
and the number of $\beta_{-,y'}$ with absolute values larger than one can be obtained from the non-Bloch band theory as $|N_y\overline{W}_x|$~\cite{zhang2020correspondence,SupMat}, 
with $\overline{W}_x$ the average spectral winding number along $x$ direction for different \blue{$\mathbf{k_y}=(k_{y_1},k_{y_2},...k_{y_{m-1}})$}, defined for a given energy $E$ as
\blue{\begin{eqnarray}
    \label{eq:winding}
    \overline{W}_x&=&\int_{\rm BZ} \frac{d\mathbf{k_y}}{(2\pi)^{m-1}} W_x(\mathbf{k_y},E)\\    
    &=&\int_{\rm BZ} \frac{d\mathbf{k_y}} {(2\pi)^{m-1}}\int_{0}^{2\pi}\frac{d k_x}{2\pi}\frac{d}{d k_x}{\rm arg}[E(k_x,\mathbf{k_y})-E],\nonumber
\end{eqnarray}
with BZ the $(m-1)$D Brillouin zone of $\mathbf{y}$.} Note that in finite systems, the integral of $\mathbf{k_y}$ becomes a summation $\sum_{\mathbf{k_y}}W_x(\mathbf{k_y})/N_y$.
Since $W_x(\mathbf{k_y},E)$ is quantized for each $k_y$, $N_y\overline{W}_x$ always take integer values, and  describes the total spectral winding number along $x$ direction.

To further deduce the conditions for skin defect states to emerge,
we rewrite the eigensolution of Eq.\eqref{eq:phi} and Eq.\eqref{eq:solution_form} as
\begin{eqnarray}
    \label{eq:state_T_lam_gam}
    \Phi_{x}=T\Lambda_{\beta}^{x}\Gamma
\end{eqnarray}
with
$\Gamma=\left[\gamma_{+,1}~ \gamma_{+,2}~\cdots~\gamma_{+,N_{y}}~\gamma_{-,1}~\cdots~\gamma_{-,N_{y}}\right]^{\rm T}$,
$\Lambda_\beta={\rm diag}\left[\beta_{+,1},\beta_{+,2},\cdots,\beta_{+,N_{y}},\beta_{-,1},\beta_{-,2},\cdots,\beta_{-,N_{y}}\right]$,
and $T=\left[I_{N_{y}\times N_{y}}~I_{N_{y}\times N_{y}}\right]$ composed by two identity matrices.
Substituting Eq.\eqref{eq:state_T_lam_gam} into the boundary conditions of Eq.\eqref{eq:transformed_boundary}, we have
\begin{eqnarray}
    \label{eq:T_boundary}
    \begin{bmatrix}
       Q\Phi_0-BQ\Phi_{N_{x}}\\
       Q\Phi_{{N_x}+1}-BQ\Phi_{1}
    \end{bmatrix}=M\Gamma=0,
\end{eqnarray}
with $M$ a $2N_{y}$-dimensional square matrix, whose elements are given by
\begin{eqnarray}
    \label{eq:M_elements1}
    M_{r,y'_{\pm}}&=&q_{r,y'}-q_{r,y'}\beta_{\pm,y'}^{N_{x}}b_r\\
    \label{eq:M_elements2}
    M_{r+N_{y},y'_{\pm}}&=&q_{r,y'}\beta_{\pm,y'}^{N_{x}+1}-q_{r,y'}\beta_{\pm,y'}b_r,
\end{eqnarray}
$q_{r,y'}$ the elements of $Q$ matrix, $y'_+=y'$, $y'_-=y'+N_{y}$, and $r,y'=1,2,...N_{y}$.

In the thermodynamic limit $N_{x}\rightarrow \infty$,
these elements can be further simplified to their leading terms of $\beta_{\pm,y'}^{N_{x}}$ with $|\beta_{\pm,y'}|>1$,
and $M_{r,y'_\pm}$ takes nonzero values only when $b_r\neq0$, i.e., when $r$ indexes $y$-positions without defects, whose number is given by $L_0=N_y-L_d$.
Keeping only the rows and columns with nonzero leading terms, we obtain
a $(N_{y}+L_0)\times N_{y}(1+|\overline{W}_x|)$ matrix $M_R$ (explicit form can be found in Supplemental Materials~\cite{SupMat}),
and skin defect states composed of $\gamma_{\pm,y'}$ with $|\beta_{\pm,y'}|>1$ is reduced to the solution of
\begin{eqnarray}
M_R\Gamma_R=0\label{eq:T_boundary_prime}
\end{eqnarray}
where $\Gamma_R$ is a column vector with $N_{y}(1+|\overline{W}_x|)$ elements.
Note that the number of 
defect states localized by NHSE along $x$ direction shall scale with the system's size $N_x$, and tends to infinity in the thermodynamic limit.
Therefore, skin defect states emerge only when
\begin{eqnarray}
N_{y}(1+|\overline{W}_x|)>{\rm rank}(M_R),\label{eq:rank1}
\end{eqnarray}
so that $M_R$ is underdetermined and Eq.~\eqref{eq:T_boundary_prime} has an infinite number of solutions.
On the other hand, the rank of $M_R$ satisfies
\begin{eqnarray}
{\rm rank}(M_R)\leqslant N_{y}+L_0.\label{eq:rank2}
\end{eqnarray}
As elements of $M_R$ are determined by $q_{r.y'}$ through Eqs.~\eqref{eq:M_elements1} and ~\eqref{eq:M_elements2}, 
the equal sign in Eq.~\eqref{eq:rank2} holds when the matrix $Q$ is a totally non-singular matrix, i.e., any minor of $Q$ is not zero. 
In this case, Eq.~\eqref{eq:rank1} can be simplified as
\begin{eqnarray}
|\overline{W}_x|>L_0/N_y.\label{eq:restriction}
\end{eqnarray}
Namely, skin defect states emerge when the spectral winding number $x$ direction ($|N_y\overline{W}_x|$) exceeds the lengths of lattices along $y$ without defect ($L_0$).
In our \blue{2D example for numerical demonstration}, the totally non-singular condition of $Q$ is satisfied when $N_y$ is a prime number~\cite{SupMat}.
This correspondence is verified by our numerical results in Fig. \ref{fig:model}(b) and (d), where the boundary between skin defect states and extended states
is seen to be marked by the critical value of the average winding number, $|\overline{W}^c_x|=L_0/N_y$.
\blue{This is in sharp contrast to topological characterization of NHSE in most contemporary studies, where skin states distinguish only the signs of topological invariants, but not their specific values.
In particular, a recent study unifies boundary and dislocation skin effects by associating their emergence to the jump of a real-space topological invariant known as the localizer index~\cite{chadha2024real}, but not to specific jump values either.
}

Before moving on to applying our results for characterizing fractal defects,
we note that 
\blue{the above derivation does not reply on the explicit form of the coupling matrix $J$, suggesting that boundary conditions and translational symmetries along $\mathbf{y}$ directions are irrelevant to our conclusion. A crucial condition is the totally non-singularity of $Q$ matrix.}
physically, it ensures that $J$ cannot be transformed into diagonalized blocks by rearranging the basis, 
i.e., the system cannot be separated into several decoupled subsystems composed by different $y'$~\cite{SupMat}.  
On the other hand, having some minors of $Q$ being zero is a rather strong condition that requires fine tuning of the hopping matrix $J$,
which is sensitive to perturbation.
In other words, even without $Q$ being totally non-singular (e.g., our \blue{2D example} with $N_y$ being a composed number, as demonstrated below),
the restriction of Eq.~\eqref{eq:restriction} on skin defect states can be recovered by introducing disorder to the hopping matrix $J$,
which is generally inevitable in actual physical systems.

\emph{Characterization of fractal defects.}---
Since the average winding number $\overline{W}_x$ is related to the defect's total length by the emergence of skin defect states,
it also provides a mean to characterize the fractal features of defects through the skin defect states.
As examples, we arrange the defects as a generalized Cantor set at the $n$th iteration,
whose fractal dimension can be given by
$$D_f=\frac{\log N_{r_n}}{\log(1/r_n)},$$
where $r_n= (1-\gamma)^n/2^n$ is the length of each defective interval at the $n$th iteration (with $N_y$ normalized to $1$),
$N_{r_n}=L_d/(N_yr_n)$ is the number of these intervals needed to cover all the defects,
and $0<\gamma<1$ is the ratio of the interval to be removed at each iteration.
Thus, together with Eq.~\eqref{eq:restriction}, 
we can establish a relation between $D_f$ and  the average winding number $\overline{W}_x$ through the emergence of skin defect states,
\begin{eqnarray}
|\overline{W}_x^c|=1-r_n^{1-D_f},\label{eq:W_Df}
\end{eqnarray}
where $\overline{W}_x^c$ is the critical value of the average winding number that marks the boundary between skin defect states and extended states.

\begin{figure}
    \includegraphics[width=0.9\linewidth]{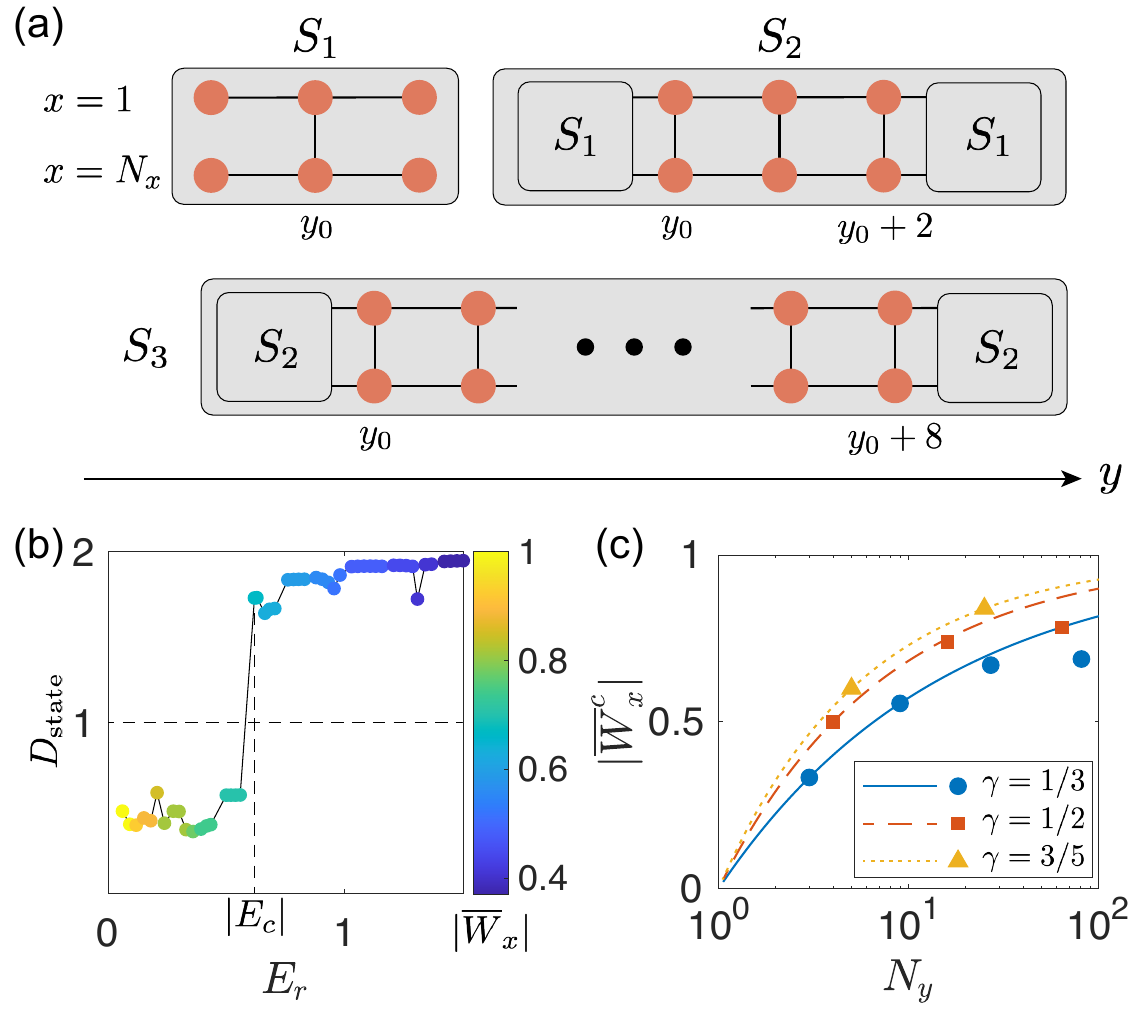}  
    \caption{Fractal defects as generalized Cantor sets with $\gamma=(m-2)/m$, and their characterization through the emergence of skin defect states.
    (a) Sketch of the minimal lattice hosting fractal defects as a Cantor set with $\gamma=1/3$, with $S_n$ represents its $n$th iteration.
    (b) Fractal dimension $D_{\rm state}$ of eigenstates with eigenenergies $E(r)\approx E_re^{i\theta}$, where $\theta=\pi/9$ is chosen. 
    Colors indicate the average winding number of the corresponding eigenenergy.
     The critical eigenenergy $E_c$ is defined as the first one with $D_{\rm state}>1$ when increasing $E_r$,
     which determine the critical value of the average winding number, $|\overline{W}_x^c|$.
    (c) $|\overline{W}_x^c|$ for defects as Cantor sets with different $\gamma$.
    Solid, dash, and dotted lines are analytical results of Eq.~\eqref{eq:W_Df}.
    Circles, squares, and triangles are obtained numerically by identifying $E_c$ for each case, with $n=1,2,...$ from left to right for each set of marks.
    The system's size is chosen as $N_x=3000$ and $N_y=m^n$, namely the minimal lattice that accommodates the Cantor set, with $m=3$ and $n=3$ for panel (b). Other parameters are the same as in Fig. \ref{fig:model}.
    A random disorder $\lambda\in[-0.2,0.2]$ is added to each hopping term, to ensure totally non-singularity of $Q$.    }
    \label{fig:fractal}
\end{figure}

As examples, we consider generalized Cantor sets with $\gamma=(m-2)/m$, $m$ being an integer larger than $2$, 
and $N_y=m^n$ the minimal lattice to support its $n$th iteration, as sketched in Fig. \ref{fig:fractal}(a) for $m=3$.
The characterization of fractal defects by skin defect states are demonstrated in Fig. \ref{fig:fractal}(b) and (c).
Explicitly, we  consider eigenenergies along a 1D trajectory in the complex energy plane, $E(r)=E_re^{i\theta}$, and calculate $D_{\rm state}$ of their corresponding eigenstates. \blue{Note that $E_r e^{i\theta}$ may not be exactly the eigenenergy of a finite-size system, and $E(r)$ is chosen as the eigenenergy closest to $E_r e^{i\theta}$ in our numerical calculation.}
Skin defect states are expected to have $D_{\rm state}<1$ as they localize non-uniformly along the 1D subsystem hosting the defects [as discussed for Fig. \ref{fig:model}(c)], while bulk states are expected to have $D_{\rm state}\approx2$.
Accordingly, we can determine $\overline{W}_x^c$, the critical value of average winding number separating these states, as shown in Fig. \ref{fig:fractal}(b).
In Fig. \ref{fig:fractal}(c), we demonstrate the numerical results of $\overline{W}_x^c$ for generalized fractal defects with different $\gamma$ and $n$,
which match well  with the analytical prediction of Eq.~\eqref{eq:W_Df} for relatively small $N_y$.
Deviation between analytical and numerical results at larger $N_y$ can be attributed to the finite-size effect along $x$ direction and can be reduced by increasing $N_x$, as discussed in Sec. II B Supplemental Materials~\cite{SupMat}.

\begin{figure}
    \includegraphics[width=1\linewidth]{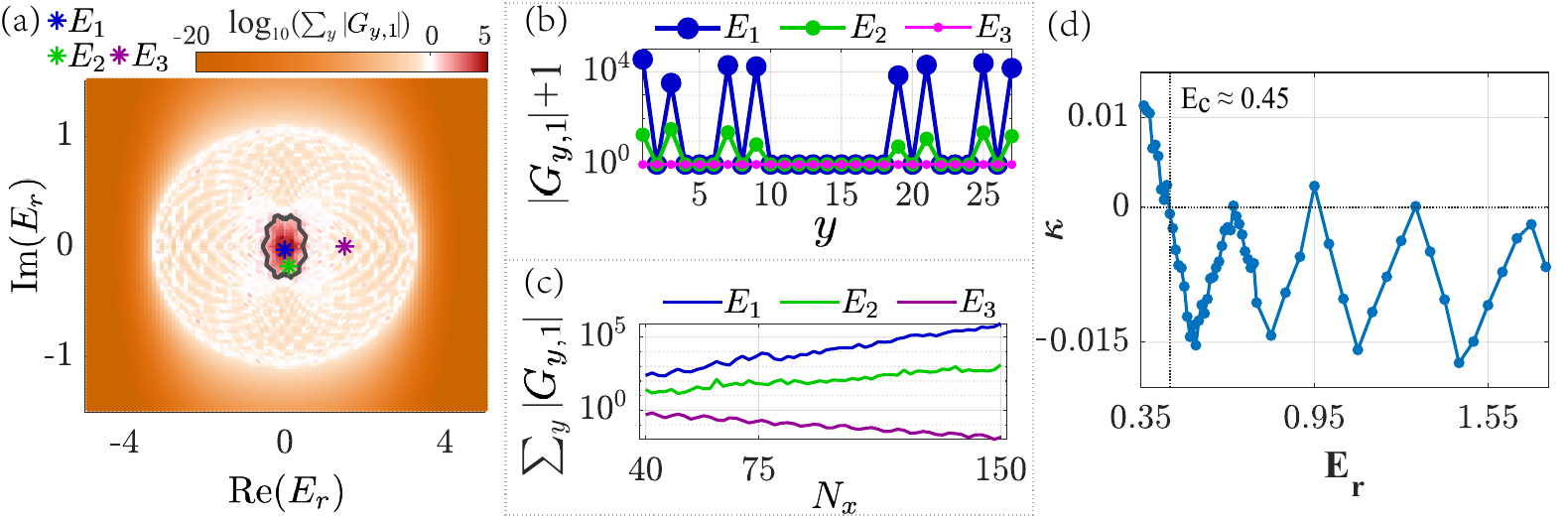}  
    \caption{(a) Summed amplification ratio at defects for different reference energy $E_r$,
    with $N_x=120$, $N_y=27$, and defects arranged as the 3rd iteration of Cantor set with $\gamma=1/3$.
    Black curve corresponds to $E_r$ with $\overline{W}_x=\overline{W}_x^c$.
    Signal amplification is seen to manifest in the region with $\overline{W}_x>\overline{W}_x^c$, with $\sum_y|G_{y,1}|\gg1$.
    (b) Amplification ratio across the defects, with (blue) $E_1=-0.0429i$, (green) $E_2=0.1429-0.1714i$, and (purple) $E_3=1.5$, marked by the stars with the same colors in (a).
    Here we consider $|G_{y,1}|+1$ instead to focus on the amplification (with $|G_{y,1}|\gg1$) in the logarithmic plot.
     Peaks corresponds to the defective lattice sites.
     (c) Scaling of the summed amplification ratio, for the same reference energies as in (b).
     (d) Slopes of the linear fitting ${\rm log}_{10}(\sum_y|G_{y,1}|)=\kappa N_x+b$, versus real reference energy $E_r$.
     The vertical dash line represents the critical energy $E_c$ read from (a) with $|\overline{W}_x|=|\overline{W}_x^c|$.
     The signal amplification is characterized by $\kappa>0$ when $E_r<E_c$.
     Other parameters are the same as in Fig. \ref{fig:model}.
     Random disorders $\lambda\in[-0.2,0.2]$ is added to the hopping terms, to ensure totally non-singularity of $Q$.
   }
    \label{fig:amplification}
\end{figure}

\emph{Amplified response of the skin defect states.}---
As a probe of the spectral winding topology and fractal structure of defects,
we consider an amplified steady-state response to external driving fields extracted from the Green's function
\begin{eqnarray}
G=\frac{1}{E_r-H},
\end{eqnarray}
where $E_r$ is a reference energy for measuring the response.
Explicitly, the response ratio at site $(N_x/2,y)$ of the defects to a distant external drive at site $(1,1)$ is given by the element
$G_{(N_x/2,y),(1,1)} \equiv G_{y,1}$~\cite{wanjura2020topological,xue2021simple}.
It is found that total response along the defects is amplified only for reference energies with $|\overline{W}_x|>|\overline{W}_x^c|$, namely the region with skin defect states, as shown in Fig. \ref{fig:amplification}(a).
Fig. \ref{fig:amplification}(b) further demonstrates the amplification ratio across the defects, which is the most pronounced at the defective sites.

A typical feature of the non-Hermitian signal amplification is its scaling with the system's size, since the signal is gradually amplified when it travels through the lattice.
As shown in Fig. \ref{fig:amplification}(c), the amplification ratio increases (decreases) with the system's size when $|\overline{W}_x|>|\overline{W}_x^c|$ ($|\overline{W}_x|<|\overline{W}_x^c|$), indicating the topological origin of the signal amplification at the defects.
In particular, the summed amplification ratio at defects is approximately $\sum_y|G_{y,1}|\approx e^{\kappa N_x}$ with $\kappa$ the slopes of the linear functions fitted to the curves in Fig. \ref{fig:amplification}(c).
In Fig. \ref{fig:amplification}(d), we illustrate $\kappa$ as a function of real reference energy $E_r$, where positive (with amplified response) and negative (without amplified response) $\kappa$ are well separated by the critical reference energy $E_c$ corresponding to $|\overline{W}_x|=|\overline{W}_x^c|$.
In addition, a strong oscillation of $\kappa$ is seen when $E_r>E_c$. 
This is because in this region the system possesses extended eigenstates and behaves as in the absence of defects, where the magnitude of the response is generally suppressed (thus a negative $\kappa$).
However, it tends to take the same size-independent magnitude as the driving field when $E_r$ approaches an eigenenergy~\cite{xue2021simple}, resulting in the peaks
with $\kappa\approx 0$.
Note that $\kappa>0$ is seen at the peak near $E_r=0.95$ in Fig.~\ref{fig:amplification}(d),
which is possibly due to numerical inaccuracy as the Green's function diverges when $E_r$ falls exactly at an eigenenergy.

\emph{Conclusions and discussion.}---
In this paper, we establish a direct correspondence between spectral winding topology, fractal characteristics, and defect-localized states for non-Hermitian lattices in arbitrary dimensions. 
\blue{Our formalism is strictly derived when the hopping matrix along one direction commute with the others; yet it can be generalized to arbitrary higher-dimensional models without such a restriction~(see Sec. V of Supplemental Materials~\cite{SupMat}),
making the results even more broadly applicable. 
The universality of our formalism is further verified by numerical results of two different scenarios with a domain wall between Hermitian and non-Hermitian and 2D block defects, respectively, as demonstrated in Sec. VI of Supplemental Materials~\cite{SupMat}. Detailed analyses of such extensions still await further exploration.}

In addition to the static eigenstates, we demonstrate that these defect-localized states manifest as amplified responses to external driving fields, which can be measured experimentally in both classical and quantum systems~\cite{wanjura2020topological,xue2021simple,li_quantized_2021},
making our finding highly relevant to various simulation platforms that realize non-Hermitian lattices \cite{Lin:21,zhang_observation_2021,PhysRevResearch.2.013280,zhang2021acoustic,wang2023non,liu2021non,helbig_generalized_2020,zou2021observation}.
Our findings not only extend the understanding of non-Hermitian topological phases but also provide a powerful approach for discovering novel defect states. 


\emph{Acknowledgement.}--- This work is supported by
National Natural Science Foundation of China (No. 12474159).

\begin{widetext}

\begin{center}
\textbf{\large Supplemental Materials}
\end{center}

\tableofcontents
\setcounter{secnumdepth}{2}

	\setcounter{equation}{0} \setcounter{figure}{0} \setcounter{table}{0} %
	\renewcommand{\theequation}{S\arabic{equation}} \renewcommand{\thefigure}{S%
		\arabic{figure}} \renewcommand{\bibnumfmt}[1]{[S#1]} 

\subsection{I. Formal solution of $H_{\rm 1D}$ in the main text}
In this section we derive the formal solution [Eq.~(7) in the main text] for the 1D model $H_{\rm 1D}$ in Eq.~(3) in the main text, with defects between sites at $x=N_x$ and $x=1$, acting as a boundary of the 1D lattice.
Its bulk eigenequation is transformed into
\begin{eqnarray}
    \label{eq:transformed_bulk_supp}
    E\Phi_x=\Lambda\Phi_x+t_R\Phi_{x-1}+t_L\Phi_{x+1}
\end{eqnarray}
by the similarly transformation of Eq.~(6) in the main text.
Here $\Lambda$ is a diagonal matrix and 
$\Phi_x=[\phi _{x,1}~\phi _{x,2}~\cdots~\phi _{x,N_y}]^{\rm T}$ is a column vector.
As the bulk equation is decoupled for different $y'$, 
the formal solution is in the same form as for the Hatano-Nelson model, which has been found in Ref. \cite{guo2021exact}.
Explicitly, we can take an ansatz $\phi_{x,y'}=\beta_{y'}^x\gamma_{y'}$ and substitute it into Eq.~\eqref{eq:transformed_bulk_supp}, which yields
\begin{eqnarray}
    \label{eq:beta_bulk}
    E=(\Lambda)_{y'}+\beta_{y'}^{-1}t_R+\beta_{y'}t_L.
\end{eqnarray}
According to the Vieta theorem, for any given eigenenergy $E$, there are two solutions $\beta_{+,y'}$ and $\beta_{-,y'}$, which satisfy:
\begin{eqnarray}
    \beta_{+,y'}\beta_{-,y'}=\frac{t_R}{t_L},
\end{eqnarray}
therefore we set the forms of $\beta_{+,y'}$ and $\beta_{-,y'}$ as:
\begin{eqnarray}
    \label{eq:beta_expanded}
    \beta_{+,y'}=Rr_{y'}e^{-i\alpha_{y'}},\nonumber\\
    \beta_{-,y'}=Rr_{y'}^{-1}e^{i\alpha_{y'}}
\end{eqnarray}
with $R=\sqrt{\frac{t_R}{t_L}}$, $r_{y'}$ and $\alpha_{y'}$ being real. 
The sign of $\alpha_{y'}$ is chosen according to i) the Fourier transformation of operators, $C_x=\frac{1}{\sqrt{N_{x}}}\sum_{k_x}e^{i k_x x}C_{k_x}$, which leads to $E=\left(\Lambda\right)_{y'}+e^{-ik_x}t_R+e^{ik_x}t_L$ under PBCs of the 1D model; and ii) we will see later that it is $\beta_{-,y'}$ which contributes to the eigensolutions when setting $\alpha_{y'}>1$ as in the main text.
Formally, the solution of correspondent eigenstate can be written as:
\begin{eqnarray}
    \label{eq:solution_form_supmat}
    \phi_{y',x}=
    \gamma_{+,y'}\beta_{+,y'}^x+\gamma_{-,y'}\beta_{-,y'}^x=
    \gamma_{+,y'}R^xr_{y'}^xe^{-ix\alpha_{y'}}+\gamma_{-,y'}R^xr_{y'}^{-x}e^{ix\alpha_{y'}},
\end{eqnarray}
where $\gamma_{+,y'}$ and $\gamma_{-,y'}$ need be determined by boundary conditions.

In the absence of any defect (i.e., when the 1D Hamiltonian is under PBCs), we have the boundary conditions $\phi_{0,y'}=\phi_{N_{x},y'}$ and $\phi_{1,y'}=\phi_{N_{x}+1,y'}$,that is:
\begin{eqnarray}
    \gamma_{+,y'}+\gamma_{-,y'}&=&\gamma_{+,y'}R^{N_{x}}r_{y'}^{N_{x}}e^{i{N_{x}}\alpha_{y'}}+\gamma_{-,y'}R^{N_{x}}r_{y'}^{-{N_{x}}}e^{-i{N_{x}}\alpha_{y'}},\label{eq:boundary1_PBC_sup}\\
    \gamma_{+,y'}Rr_{y'}e^{i\alpha_{y'}}+\gamma_{-,y'}Rr_{y'}^{-1}e^{-i\alpha_{y'}}&=&\gamma_{+,y'}R^{N_{x}+1}r_{y'}^{N_{x}+1}e^{i{N_x+1)}\alpha_{y'}}+\gamma_{-,y'}R^{N_{x}+1}r_{y'}^{-{(N_{x}+1)}}e^{-i{(N_{x}+1)}\alpha_{y'}},\label{eq:boundary2_PBC_sup}
\end{eqnarray}
which lead to $r_y'=R$, $\gamma_{+,y'}=0$ and $\alpha_{y'}=2\pi l_x/N_{x}$ in Eq.~\eqref{eq:solution_form_supmat}, with $l_x=1,2,..., N_{x}$. (notice that we have set $R>1$ and $r_{y'}>1$ as in the main text, without loss of generality).
Substituting the solution of $\phi_{x,y'}$ into Eq.~\eqref{eq:beta_bulk} and Eq.~\eqref{eq:beta_expanded},
the PBC spectrum are thus given by $N_{y}$ elliptises centering at the diagonal elements of $\Lambda$,
\begin{eqnarray}
    \label{eq:model2_PBC_spectrum}
    E_{y'}(k_x)=(\Lambda)_{y'}+e^{-ik_x}t_R+e^{ik_x}t_L=(\Lambda)_{y'} +\sqrt{t_Rt_L}[(R+R^{-1}){\rm cos}k_x+i(R^{-1}-R){\rm sin}k_x],
\end{eqnarray}
with $\alpha_{y'}$ replaced by the crystal momentum $k_x$ under PBCs.


\subsection{II. Deduction of skin defect states}\label{sec:deduction}
\subsubsection{A. Emergence of skin defect states in the thermodynamic limit}
In this section we provide further details of our derivation of the skin defect states.
We begin with Eqs.~(10) to (12) in the matin text:
\begin{eqnarray}
    \label{eq:T_boundary_supmat}
    \begin{bmatrix}
       Q\Phi_0-BQ\Phi_{N_{x}}\\
       Q\Phi_{N_{x}+1}-BQ\Phi_{1}
    \end{bmatrix}&=&M\Gamma=0,\\
    \label{eq:M_elements1_supmat}
    M_{r,y'_{\pm}}&=&q_{r,y'}-q_{r,y'}\beta_{\pm,y'}^{N_{x}}b_r,\\
    \label{eq:M_elements2_supmat}
    M_{r+N_{y},y'_{\pm}}&=&q_{r,y'}\beta_{\pm,y'}^{N_{x}+1}-q_{r,y'}\beta_{\pm,y'}b_r,
\end{eqnarray}
with $q_{r,y'}$ the elements of $Q$ matrix, $y'_+=y'$, $y'_-=y'+N_{y}$, and $r,y'=1,2,...N_{y}$.
Here $\Gamma=\left[\gamma_{+,1}~ \gamma_{+1,2}~\cdots~\gamma_{+,N_{y}}~\gamma_{-,1}~\cdots~\gamma_{-,N_{y}}\right]^{\rm T}$ is a column matrix corresponding to the eigenstates in Eqs.~(6) and (7) in the main text.
Thus, by taking $\beta_{\pm,y'}$ as variables, 
their nontrivial solutions are constrained by Eq.~\eqref{eq:T_boundary_supmat}, which requires 
\begin{eqnarray}
{\rm det}[M]=0.\label{eq:det_M}
\end{eqnarray}

Next, before analyzing solutions of skin states localized at defects, we first unveil how Eq.~\eqref{eq:T_boundary_supmat} is related to bulk eigenstates in the absence of defects, or in other words, when the 1D system is under PBCs.
Without defects, different components of $y'$ can be decoupled not only in the bulk, but also at the boundary, since now $B$ is an identity matrix. Therefore, Eq.~\eqref{eq:T_boundary_supmat} is decoupled into $M_{y'}\Gamma_{y'}=0$ with $M_{y'}$ a $2\times 2$ matrix for the $y'$th component, where the power of $\beta_{\pm,y'}$ is on the order of $N_{x}$.
Taking $\beta_{\pm,y'}$ as two variables, it gives Eqs.~\eqref{eq:boundary1_PBC_sup} and \eqref{eq:boundary2_PBC_sup} in the last section, which uniquely determine $|\beta_{-,y'}|^{N_{x}}=1$ (and $|\beta_{+,y'}|=\frac{t_R}{t_L}$, which is irrelevant to the eigenstates).
Therefore, different eigenstates are given by different phase factor ($\alpha_{y'}$) of $\beta_{\pm,y'}$, and their number tends to infinity in the thermodynamic limit $N_{x}\rightarrow \infty$.

In the presence of defects (i.e., removing some hoppings between $x=N_x$ and $x=1$), different components of $y'$ are mixed with each other, and we need to examine the matrix $M$ as a whole in Eq.~\eqref{eq:det_M}, where the powers of variables $\beta_{\pm,y'}$ are on the order of $N_{x}$.
Thus, directly analyzing Eq.~\eqref{eq:det_M} may encounter solutions with finite $|\beta_{\pm,y'}^{N_{x}}|$, which indicate
$|\beta_{\pm,y'}|\rightarrow 1$ in the thermodynamic and thus do not represent skin states.

To determine the emergence of skin defect states in the thermodynamic limit, we need to find solutions with $|\beta_{\pm,y'}|\neq 1$.
To this end, we simplify Eq.\eqref{eq:M_elements1_supmat} and Eq.\eqref{eq:M_elements2_supmat} as followed:
\begin{eqnarray}
    \label{eq:M_elements1_simplify_supmat}
    M_{r,y'_{\pm}}&=&q_{r,y'}-q_{r,y'}\beta_{\pm,y'}^{N_{x}}b_r\\
    &\underset{N_{x}\rightarrow \infty}{=}&\begin{cases}
        -q_{r,y'}\beta^{N_{x}}_{\pm,y'}b_r,\quad {\rm when}~|\beta_{\pm,y'}|>1;\nonumber\\
       q_{r,y'},\qquad\qquad\ \ \, {\rm when}~|\beta_{\pm,y'}|<1;
       \end{cases}\\
    \label{eq:M_elements2_simplify_supmat}
    M_{r+N_{x},y'_{\pm}}&=&q_{r,y'}\beta_{\pm,y'}^{N_{x}+1}-q_{r,y'}\beta_{\pm,y'}b_r\\
    &\underset{N_{y'}\rightarrow \infty}{=}&\begin{cases}
        q_{r,y'}\beta_{\pm,y'}^{N_{x}+1},\quad\quad {\rm when}~|\beta_{\pm,y'}|>1;\nonumber\\
       -q_{r,y'}\beta_{\pm,y'}b_r,\quad {\rm when}~|\beta_{\pm,y'}|<1.
       \end{cases}
\end{eqnarray}
Note that here we have assumed $|\beta_{\pm,y'}|\neq 1$ for simplicity. As we shall see later, having $|\beta_{\pm,y'}|=1$ for some values of $y'$ does not affect our deduction.

Now we consider eigenenergies with the same winding number $W_x$. Following our discussion in the main text,
the number of $\beta_{-,y'}$s in their corresponding eigensolutions with $\beta_{-,y'}>1$ is $|W_x|$.
We label them as
$\beta_{-,w_1},\beta_{-,w_2},\cdots,\beta_{-,w_{|W_x|}}$, and the rest $(N_{y}-|W_x|)$ $\beta_{-,y'}$s satisfying $|\beta_{-,y'}|\leqslant1$ as $\beta_{-,\overline{w}_1},\beta_{-,\overline{w}_2},\cdots,\beta_{-,\overline{w}_{N_{y}-|W_x|}}$.
Next, we consider the submatrix $M_R$ formed by columns of $M$ with $|\beta_{\pm,y'}|>1$, namely, $\beta_{+,y'}$ with $y'=1,2,...,N_{y}$ and $\beta_{-,w_1},\beta_{-,w_2},\cdots,\beta_{-,w_{|W_x|}}$. Its explicit form, after removing the exponential terms of $\beta_{\pm,y'}^{N_{x}}$, is given by
          
    \begin{eqnarray}
        \label{eq:M_R}
        &\widetilde{M}_R=M_R\Lambda_{\beta_R}^{-N_{x}}=\nonumber\\
      &\left[\begin{array}{cccc|cccc}
        -q_{1,1}b_{1}&-q_{1,2}b_{1}  &\cdots  &-q_{1,N_{y}}b_{1}  &-q_{1,w_1}b_{1}  &-q_{1,w_2}b_{1}  &\cdots  &-q_{1,w_{|W_x|}}b_{1} \\
    
        -q_{2,1}b_{2}&-q_{2,2}b_{2}  &\cdots  &-q_{2,N_{y}}b_{2}  &-q_{2,w_1}b_{2}  &-q_{2,w_2}b_{2}  &\cdots  &-q_{2,w_{|W_x|}}b_{2} \\
    
        \vdots&\vdots  &\vdots  &\vdots  &\vdots  &\vdots  &\vdots  &\vdots \\
    
        -q_{N_{y},1}b_{N_{y}}&-q_{N_{y},2}b_{N_{y}}  &\cdots  &-q_{N_{y},N_{y}}b_{N_{y}}  &-q_{N_{y},w_1}b_{N_{y}}  &-q_{N_{y},w_2}b_{N_{y}}  &\cdots  &-q_{N_{y},w_{|W_x|}}b_{N_{y}} \\

            \hline
    
        q_{1,1}\beta_{+,1}&q_{1,2}\beta_{+,2}  &\cdots  &q_{1,N_{y}}\beta_{+,N_{y}}  &q_{1,w_1}\beta_{-,w_1}  &q_{1,w_2}\beta_{-,w_2}  &\cdots  &q_{1,w_{|W_x|}}\beta_{-,w_{|W_x|}} \\
    
        q_{2,1}\beta_{+,1}&q_{2,2}\beta_{+,2}  &\cdots  &q_{2,N_{y}}\beta_{+,N_{y}}  &q_{2,w_1}\beta_{-,w_1}  &q_{2,w_2}\beta_{-,w_2}  &\cdots  &q_{2,w_{|W_x|}}\beta_{-,w_{|W_x|}} \\
    
        \vdots&\vdots  &\vdots  &\vdots  &\vdots  &\vdots  &\vdots  &\vdots \\
    
         q_{N_{y},1}\beta_{+,1}&q_{N_{y},2}\beta_{+,2}  &\cdots  &q_{N_{y},N_{y}}\beta_{+,N_{y}}  &q_{N_{y},w_1}\beta_{-,w_1} &q_{N_{y},w_2}\beta_{-,w_2}  &\cdots  &q_{N_{y},w_{|W_x|}}\beta_{-,w_{|W_x|}} \\
     
      \end{array}\right]_{2N_{y}\times(N_{y}+|W_x|)}
      \nonumber\\
    \end{eqnarray}
   with a diagonal matrix
    \begin{eqnarray}
        \Lambda_{\beta_R}=\nonumber{\rm diag}(\beta_{+,1},\beta_{+,2},\cdots,\beta_{+,N_{y}},\beta_{-,w_1},\beta_{-,w_2},\cdots,\beta_{-,w_{|W_x|}}).
    \end{eqnarray}
    We also label the submatrix formed by the rest columns of $M$ corresponding to $\beta_{-,\overline{w}_1},\beta_{-,\overline{w}_2},\cdots,\beta_{-,\overline{w}_{N_{y}-|W_x|}}$ as $M_L$.
    Thus, the condition to have nontrivial eigensolutions of Eq.\eqref{eq:T_boundary_supmat}, ${\rm det}[M]=0$, is equivalent to
    \begin{eqnarray}
        \label{eq:det_Mtilde}
        {\rm det}\begin{bmatrix}
            \widetilde{M}_R&M_L
                    \end{bmatrix}=0,
    \end{eqnarray}
    since the matrix $\begin{bmatrix}\widetilde{M}_R&M_L\end{bmatrix}$ is obtained from $M$ through elementary operations (exchainging different columns and multiplication of columns by non-zero numbers).
    
Note that each $\beta_{\pm, y'}$ in Eq.~\eqref{eq:det_Mtilde} has its power on the order of $1$.
Therefore, Eq.\eqref{eq:det_Mtilde} can be considered as a finite-order ($N_{y}$-order, independent from $N_{x}$) function of $\beta_{\pm,y'}$s, and may only give a finite number of solutions that correspond to localized eigenstates.
However, these solutions do not represent skin states either, whose number shall tends to infinity in the thermodynamic limit.

To find the solutions of skin states, we note that the defects give further restrictions to the matrix $\begin{bmatrix}\widetilde{M}_R&M_L\end{bmatrix}$ by having $b_r=0$ for $L_d$ (the total length of defects) different values of $r$.
Therefore, as can be seen from Eq.~\eqref{eq:M_R}, there are only $L_0=N_y-L_d$ rows in the top half of $\widetilde{M}_R$ with non-zero elements, so that
    \begin{eqnarray}
        \label{eq:ineq_2}
        {\rm rank}(\widetilde{M}_R)\leq N_{y}+L_0.
    \end{eqnarray}
On the other hand,  Eq.\eqref{eq:det_Mtilde} can be automatically satisfied when the rank of $\widetilde{M}_R$ is smaller than the number of its column, 
    \begin{eqnarray}
        \label{eq:ineq_1}
        {\rm rank}(\widetilde{M}_R)< N_{y}+|W_x|,
    \end{eqnarray}
which allows for an infinite number of solutions of $\beta_{\pm,y'}$.
From Eq.~\eqref{eq:ineq_2}, we can see that Eq.~\eqref{eq:ineq_1} is always satisfied as along as $|W_x|>L_0$,
or, \begin{eqnarray}
|\overline{W}_x|>L_0/N_y,\label{sup_eq:Wx_average}
\end{eqnarray}
where $\overline{W}_x$ is the spectral winding number averaged over different crystal momenta $k_y$, as defined in Eq.~(8) in the main text.
In order words, in the thermodynamic limit, $|\overline{W}_x|>L_0/N_y$ ensures that there are an infinite number of solutions of $\beta_{\pm,y'}$ that give different eigenstates localized at the defects by NHSE. Furthermore, they are also the solutions of $M_R\Gamma_R=0$, making the corresponding skin states exponentially decay from $x=N_x$ to $x=1$ (with $|\beta|>1$).

Similarly, as $M_L$ has $N_{y}-|W_x|$ columns, we may obtain an infinite number of solutions when 
    \begin{eqnarray}
        \label{eq:ineq_ML}
        {\rm rank}(M_L)< N_{y}-|W_x|,
    \end{eqnarray}
    which correspond to skin defect states localize toward $x=1$.
Provided $|\beta_{\pm,y'}|\neq 1$ for each $y'$, the explicit form of $M_L$ is given by
\begin{eqnarray}
        \label{eq:M_L}
        M_L=\left[\begin{array}{cccc}
            q_{1,\overline{w}_1}&q_{1,\overline{w}_2}  &\cdots  &q_{1,\overline{w}_{N_{y}-|W_x|}} \\
    
            q_{2,\overline{w}_1}&q_{2,\overline{w}_2}  &\cdots  &q_{2,\overline{w}_{N_{y}-|W_x|}} \\
    
            \vdots&\vdots  &\vdots  &\vdots \\
    
            q_{N_{y},\overline{w}_1}&q_{N_{y},\overline{w}_2}  &\cdots  &q_{N_{y},\overline{w}_{N_{y}-|W_x|}} \\
    
        \hline
    
            -q_{1,\overline{w}_1}\beta_{-,\overline{w}_1}b_1&-q_{1,\overline{w}_2}\beta_{-,\overline{w}_2}b_1  &\cdots  &-q_{1,\overline{w}_{N_{y}-|W_x|}}\beta_{-,\overline{w}_{N_{y}-|W_x|}}b_1 \\
    
           -q_{2,\overline{w}_1}\beta_{-,\overline{w}_1}b_2&-q_{2,\overline{w}_2}\beta_{-,\overline{w}_2}b_2  &\cdots  &-q_{1,\overline{w}_{N_{y}-|W_x|}}\beta_{-,\overline{w}_{N_{y}-|W_x|}}b_2 \\
    
          \vdots&\vdots&\vdots&\vdots\\
    
           -q_{N_{y},\overline{w}_1}\beta_{-,\overline{w}_1}b_{N_{y}}&-q_{{N_{y}},\overline{w}_2}\beta_{-,\overline{w}_2}b_{N_{y}}  &\cdots  &-q_{{N_{y}},\overline{w}_{N_{y}-|W_x|}}\beta_{-,\overline{w}_{N_{y}-|W_x|}}b_{N_{y}} \\
        \end{array}\right]_{2N_{y}\times(N_{y}-|W_x|)}.
    \end{eqnarray}
We can see that it contains at least $N_{y}$ rows (the top half) with non-zero elements, thus Eq.~\eqref{eq:ineq_ML} can never be satisfied. 
Having $|\beta_{\pm,y'}|=1$ for certain $y'$ further decreases the number of columns in $M_L$ relevant to left-localized skin states (with $|\beta_{\pm,y'}|<1$), and thus decreases the value on the right-hand side of the inequality~\eqref{eq:ineq_ML}.
Therefore there is no any left-localized skin states in our model within the chosen parameter regime ($t_R>t_L$). 
In the other parameter regime with $t_L>t_R$, the system shall support left-localized skin defect states when $|\overline{W}_x|>L_0/N_y$, as it can be mapped to the above case by spatial inversion.


Note that above correspondence between skin states and the spectral winding number strictly holds when Eq.~\eqref{eq:ineq_2} takes equal sign, which requires any minor of $Q$ is non-zero (the totally non-singular condition). In the next section we will discuss the physical relevance of this condition, and extend our analysis to cases where it no longer holds.
    
\subsubsection{B. Finite-size effect of eigensolutions}
In the above deduction we have assumed the thermodynamic limit $N_x\rightarrow \infty$ in Eq.~\eqref{eq:M_elements1_simplify_supmat} to determine the skin defect states.
In our model, two types of finite-size effects may affect the localization properties of bulk states and defect states, respectively.

First, given that defects can be viewed as certain local impurity for the lattice under periodic boundary conditions along $x$, a finite-size system may also support the scale-free localization~\cite{li2021impurity}.
Consequently, for a finite $N_x$, some eigenstates with $\overline{W}_x<L_0/N_y$, which are expected to be extended bulk states following our formalism, may also become localized due to the scale-free localization, as shown in Fig.~\ref{fig:scaling}(a). 
However, we note that the scale-free states has their localization lengths proportional to $N_x$, and are eventually delocalized when $N_x\rightarrow \infty$. 
Numerically, we can see that the localized states (with the effective dimension of eigenstates $D_{\rm state}<1$) with $\overline{W}_x<L_0/N_y$ gradually become two-dimensional extended states (with their $D_{\rm state}$ tends to two) when increasing $N_x$, as shown in Fig.~\ref{fig:scaling}(a) to (c).
In Fig. \ref{fig:Dstate_size}(a) to (d), we plot $D_{\rm state}$ as a function of eigenenergies along a 1D line in the complex energy plane, $E(r)\approx E_re^{i\theta}$ with $\theta=\pi/9$, for several different values of $N_x$.  
It is seen that  for smaller $N_x$, $D_{\rm state}$ varies almost continuously, making it difficult to distinguish eigenstates with different localization and extended properties.

    \begin{figure}
        \includegraphics[width=1\textwidth]{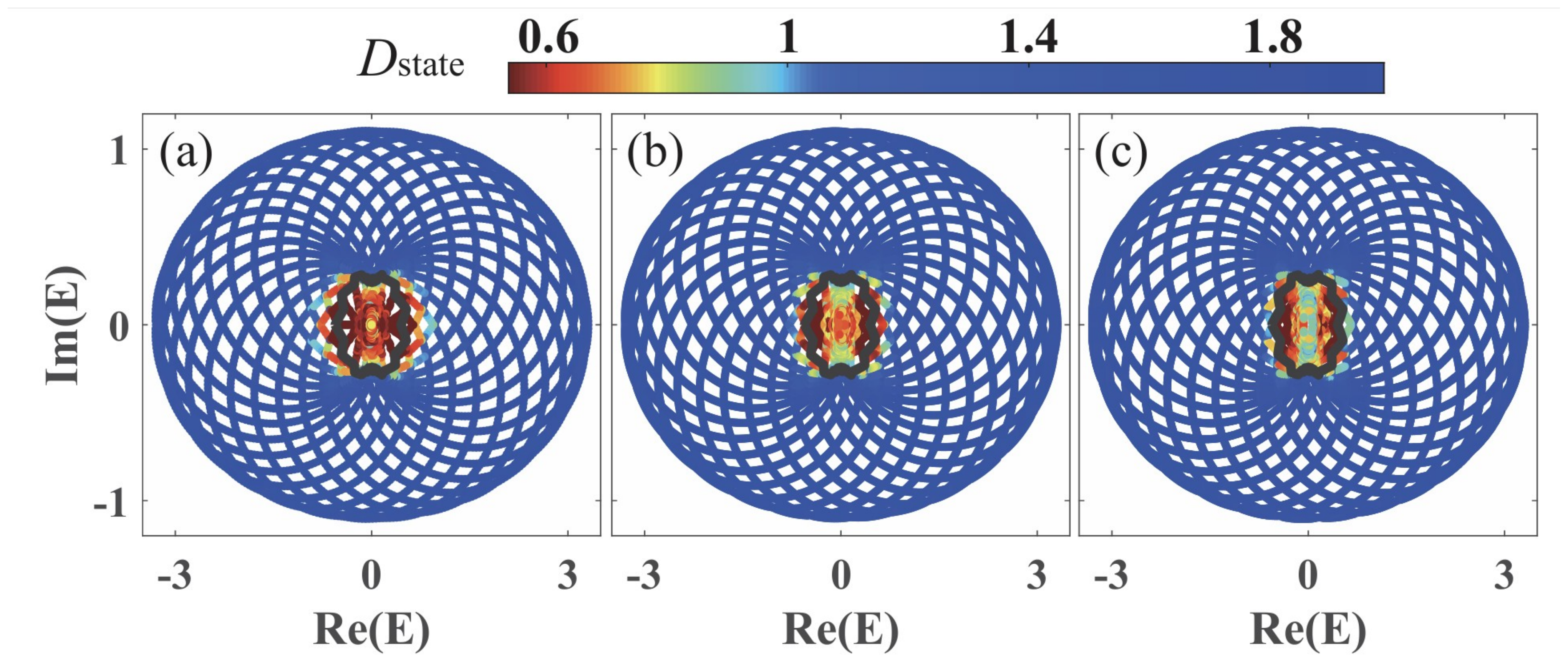}  
        \caption{Spectra of the model with (a) $N_x=200$, (b) $N_x=500$, and (c) $N_x=900$, marked by the effective dimension $D_{\rm state}$ of the eigenstate for each eigenenergy.
        For all three panels, defects are arranged as the 3rd iteration of Cantor set with $\gamma = 1/3$ [namely, as $S3$ in Fig. 2(a) in the main text]. Black loop divides the two regions corresponding to $\overline{W_x}>L_0/N_y$ (inside the loop) and $\overline{W_x}\leq L_0/N_y$ (outside the loop), respectively. 
        Eigenenergies close to $\overline{W_x}=L_0/N_y$ from outside are seen to correspond to localized states ($D_{\rm state}<1$) for small $N_x$, and become extended states ($D_{\rm state}\rightarrow 2$) when increasing the size,  showing the characteristic of scale-free localization~\cite{li2021impurity}.
        Other parameters are the same as in Fig. 1 in the main text. A random disorder $\lambda\in[-0.2; 0.2]$ is added to each
        hopping term, to ensure totally non-singularity of $Q$. }
        \label{fig:scaling}
    \end{figure}

       \begin{figure}
        \includegraphics[width=1\textwidth]{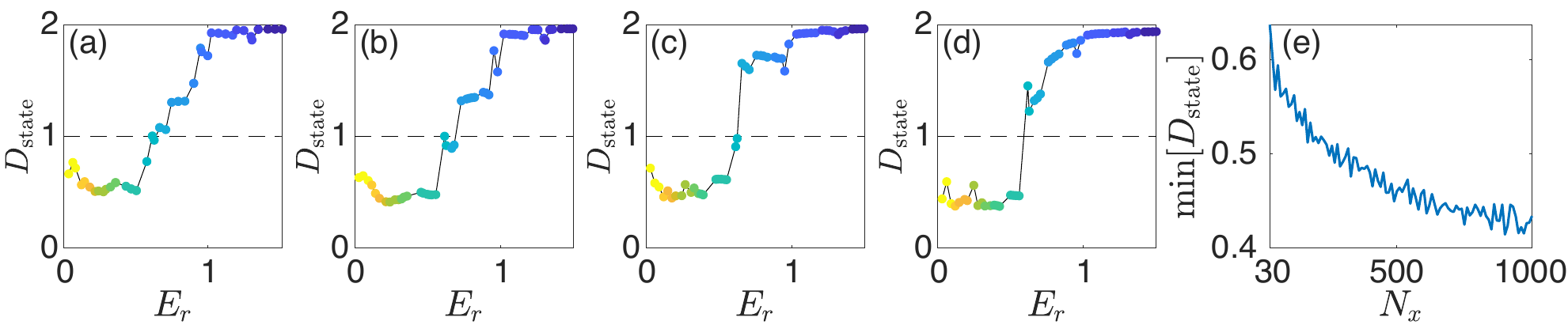}  
        \caption{
        (a) to (d) Fractal dimension $D_{\rm state}$ of eigenstates with eigenenergies $E(r)\approx E_re^{i\theta}$ and $\theta=\pi/9$, for
        different $N_x=200,500,1000,2000$, respectively.
         Colors indicate the average winding number of the corresponding eigenenergy.
         (e) The minimal value of $D_{\rm state}$ for $E(r)\approx E_re^{i\theta}$, versus the size of the system along $x$ direction. 
         Other parameters in all panels are the same as in Fig. 1 in the main text. A random disorder $\lambda\in[-0.2; 0.2]$ is added to each
        hopping term, to ensure totally non-singularity of $Q$. (a) to (d) are obtained from a single choice of the random disorder, and (e) is obtained from the average of $50$ different random disorders.
          }
        \label{fig:Dstate_size}
    \end{figure}
    
Another finite-size effect comes from the distribution properties of the skin defect states. Normally, these states distribute along the 1D defects, seemingly corresponding to an effective dimension $D_{\rm state}\approx 1$. 
On the other hand, in our derivation of the topological correspondence of skin defect states,
we have assumed thermodynamic limit along $x$, while $N_y$ needs to be finite as it determines the critical value of the spectral winding number.
In such a scenario, the skin defect states become more like 0D states when $N_x$ tends to infinity and $N_y$ is kept finite.
As a result, we shall have $D_{\rm state}$ for skin defect states drops below $1$ and keeps decreasing when increasing $N_x$, making it easier to distinguish them from bulk states for larger $N_x$.
As shown in FIg. \ref{fig:Dstate_size}(e), the minimal value of $D_{\rm state}$ (for eigeneneriges along a 1D line in the complex plane) decreases when $N_x$ increases,  and is always below $1$ in the parameter space we consider, which verifies the above conjecture.

     \subsection{III. Physical relevance of the total non-singularity}
    In previous sections, we have established connections between line defects, spectral winding numbers, the emergence of skin defect states, and the $Q$ matrix that diagonalize the coupling matrix $J$.
    In particular, skin defect states emergence when and only when $|\overline{W}_x|>L_0/N_y$, provided any minor of $Q$ is nonzero, i.e., $Q$ is a totally non-singular matrix.
    In this section, we briefly discuss two physical scenarios that are relevant to the non-singularity of $Q$.
    
    \subsubsection{A. A prime number of lattice sites along $y$ direction with translational symmetry}
    The first scenario is when the number of lattice sites along $y$ direction, $N_y$,  is a prime number.
    For translational symmetric lattices,
    the coupling matrix $J$ satisfies 
   \begin{eqnarray}
        \label{eq:sym}
        T_r^{-1}JT_r=J,
    \end{eqnarray}
    where $T_r$ is the translation matrix, with $(T_r)_{y+1,y}=1$, $y=1,2,\cdots,N_{y}-1$ and $(T_r)_{1,N_{y}}=1$, and all other elements being zero. 
    The eigenvectors of $T_r$ are $\begin{bmatrix}e^{(2\pi iy/N_{y})\times1}&e^{(2\pi iy/N_{y})\times2}&\cdots&e^{(2\pi iy/N_{y})\times N_{y}}\end{bmatrix}^{\rm T}$, where $y=1,2,\cdots,N_{y}$, 
    which are also eigenvectors of $J$ since there is no degeneracy between the eigenvalues of $T_r$ (which are $2\cos \frac{2\pi y}{N_{y}}$). 
    Therefore $Q$ can be expanded as (since it is the matrix composed by eigenvectors of $J$):
    \begin{eqnarray}
        \label{eq:Q_period}
        (Q)_{y_1,y_2}=e^{y_1 y_2(2 \pi i/N_{y})},
    \end{eqnarray}
    which is totally non-singular when $N_{y}$ is a prime number \cite{demianowicz2022universal}.

\subsubsection{B. Fully coupled components}
Physically, non-singularity of $Q$ also requires lattice sites at different $y$ to be fully coupled to each other, i.e., they cannot be separated into several decoupled subsystems.
Otherwise, we may rearrange the $y$ coordinate in the Hamiltonian, $y=(1,2,...,N_y)\rightarrow (\bar{y}_1,\bar{y}_2,...,\bar{y}_{N_y})$ with $\bar{y}_{n}=1,2,...,N_y$, so to have $J$ is a block-diagonalized matrix, with different diagonal blocks represent different separated subsystems. 
Thus the matrix $Q$, which is the eigenvector matrix of $J$, also becomes block-diagonalized and hence has zero minors.
Note that the rearrangement of components only represents how we label the lattice sites, and shall not affect the eigensolutions of the system.
In particular, it only exchange different rows and columns of $J$, which does not affect the total non-singularity of $Q$.

    \subsection{IV. Beyond the totally non-singular condition}\label{sec:beyond}
    The correspondence between skin defect states and the average spectral winding number $|W_x|$
    demands that the coupling matrix $J$ can be diagonalized with $Q^{-1}$, with $Q$ being a totally non-singular matrix (that is, any minor of $Q$ is nonzero).
    This condition ensures that Eq.~\eqref{eq:ineq_2} takes equal sign and Eq.~\eqref{eq:ineq_1} is equivalent to $L_0<|W_x|$.
    Nevertheless, even when it is not satisfied (while we still assume that $J$ can be diagonalized with $Q^{-1}$), 
    a generalized correspondence between $|W_x|$, spatial configuration of defects, zero minors of $Q$, and the occurrence of skin defect states can still be established, as we will elaborate below.

    As discussed in previous sections and the matin text, the full-PBC spectrum of the system without any defect forms $N_{y}$ ellipses in the complex energy plane, each corresponding to a component of $\phi_{x,y'}$.
    Consider an energy $E$ enclosed by $|W_x|$ of these ellipses.
    We label them as $(w_1, w_2, \cdots, w_{|W_x|})$, and the $y$-lattices without defects as $(p_1, p_2, \cdots, p_{|W_x|})$.
    Then we remove the rows in $\widetilde{M}_R$ of Eq.~\eqref{eq:M_R} with zero-value elements only (i.e., those with $b_r=0$), and obtain
    \begin{eqnarray}\label{eq:MR1}
      &\widetilde{M}_R\rightarrow\widetilde{M}_R^1=\nonumber\\
      &\left[\begin{array}{cccc|cccc}
    
        -q_{p_1,1}&-q_{p_1,2} &\cdots  &-q_{p_1,N_{y}}  &-q_{p_1,w_1}  &-q_{p_1,w_2}  &\cdots  &-q_{p_1,w_{|W_x|}}
         \\

         -q_{p_2,1}&-q_{p_2,2} &\cdots  &-q_{p_2,N_{y}}  &-q_{p_2,w_1}  &-q_{p_2,w_2}  &\cdots  &-q_{p_2,w_{|W_x|}}
         \\

        \vdots&\vdots  &\vdots  &\vdots  &\vdots  &\vdots  &\vdots  &\vdots \\

         -q_{p_{L_0},1}&-q_{p_{L_0},2} &\cdots  &-q_{p_{L_0},N_{y}}  &-q_{p_{L_0},w_1}  &-q_{p_{L_0},w_2}  &\cdots  &-q_{p_{L_0},w_{|W_x|}}
         \\

            \hline
    
        q_{1,1}\beta_{+,1}&q_{1,2}\beta_{+,2}  &\cdots  &q_{1,N_{y}}\beta_{+,N_{y}}  &q_{1,w_1}\beta_{-,w_1}  &q_{1,w_2}\beta_{-,w_2}  &\cdots  &q_{1,w_{|W_x|}}\beta_{-,w_{|W_x|}} \\
    
        q_{2,1}\beta_{+,1}&q_{2,2}\beta_{+,2}  &\cdots  &q_{2,N_{y}}\beta_{+,N_{y}}  &q_{2,w_1}\beta_{-,w_1}  &q_{2,w_2}\beta_{-,w_2}  &\cdots  &q_{2,w_{|W_x|}}\beta_{-,w_{|W_x|}} \\
    
        \vdots&\vdots  &\vdots  &\vdots  &\vdots  &\vdots  &\vdots  &\vdots \\
    
         q_{N_{y},1}\beta_{+,1}&q_{N_{y},2}\beta_{+,2}  &\cdots  &q_{N_{y},N_{y}}\beta_{+,N_{y}}  &q_{N_{y},w_1}\beta_{-,w_1} &q_{N_{y},w_2}\beta_{-,w_2}  &\cdots  &q_{N_{y},w_{|W_x|}}\beta_{-,w_{|W_x|}} \\
     
      \end{array}\right]_{(N_{y}+L_0)\times(N_{y}+|W_x|)},
      \nonumber\\
    \end{eqnarray}
Next, we multiply the $N_{y}$ columns on the left by $f_{y'}=\frac{\beta_{-,y'}}{\beta_{+,y'}}$, which gives
 \begin{eqnarray}
      &\widetilde{M}_R^1 \rightarrow \widetilde{M}_R^2=\nonumber\\
      &\left[\begin{array}{cccc|cccc}
    
        -q_{p_1,1}f_{1}&-q_{p_1,2}f_{2} &\cdots  &-q_{p_1,N_{y}}f_{N_{y}}  &-q_{p_1,w_1}  &-q_{p_1,w_2}  &\cdots  &-q_{p_1,w_{|W_x|}}
         \\

         -q_{p_2,1}f_{1}&-q_{p_2,2}f_{2} &\cdots  &-q_{p_2,N_{y}}f_{N_{y}}  &-q_{p_2,w_1}  &-q_{p_2,w_2}  &\cdots  &-q_{p_2,w_{|W_x|}}
         \\

        \vdots&\vdots  &\vdots  &\vdots  &\vdots  &\vdots  &\vdots  &\vdots \\

         -q_{p_{L_0},1}f_{1}&-q_{p_{L_0},2}f_{2} &\cdots  &-q_{p_{L_0},N_{y}}f_{N_{y}}  &-q_{p_{L_0},w_1}  &-q_{p_{L_0},w_2}  &\cdots  &-q_{p_{L_0},w_{|W_x|}}
         \\

            \hline
    
        q_{1,1}\beta_{-,1}&q_{1,2}\beta_{-,2}  &\cdots  &q_{1,N_{y}}\beta_{-,N_{y}}  &q_{1,w_1}\beta_{-,w_1}  &q_{1,w_2}\beta_{-,w_2}  &\cdots  &q_{1,w_{|W_x|}}\beta_{-,w_{|W_x|}} \\
    
        q_{2,1}\beta_{-,1}&q_{2,2}\beta_{-,2}  &\cdots  &q_{2,N_{y}}\beta_{-,N_{y}}  &q_{2,w_1}\beta_{-,w_1}  &q_{2,w_2}\beta_{-,w_2}  &\cdots  &q_{2,w_{|W_x|}}\beta_{-,w_{|W_x|}} \\
    
        \vdots&\vdots  &\vdots  &\vdots  &\vdots  &\vdots  &\vdots  &\vdots \\
    
         q_{N_{y},1}\beta_{-,1}&q_{N_{y},2}\beta_{-,2}  &\cdots  &q_{N_{y},N_{y}}\beta_{-,N_{y}}  &q_{N_{y},w_1}\beta_{-,w_1} &q_{N_{y},w_2}\beta_{-,w_2}  &\cdots  &q_{N_{y},w_{|W_x|}}\beta_{-,w_{|W_x|}} \\
     
      \end{array}\right];
      \nonumber\\
    \end{eqnarray}
  and subtract each of them from the column on the right with the same $y'$, which gives
   \begin{eqnarray}
      &\widetilde{M}_R^2 \rightarrow \widetilde{M}_R^3=\nonumber\\
      &\left[\begin{array}{cccc|cccc}
    
        -q_{p_1,1}f_{1}&-q_{p_1,2}f_{2} &\cdots  &-q_{p_1,N_{y}}f_{N_{y}}  &-q_{p_1,w_1}(1-f_{w_1})  &-q_{p_1,w_2}(1-f_{w_2})  &\cdots  &-q_{p_1,w_{|W_x|}}(1-f_{w_{|W_x|}}) 
         \\

         -q_{p_2,1}f_{1}&-q_{p_2,2}f_{2} &\cdots  &-q_{p_2,N_{y}}f_{N_{y}}  &-q_{p_2,w_1}(1-f_{w_1})  &-q_{p_2,w_2}(1-f_{w_2})  &\cdots  &-q_{p_2,w_{|W_x|}}(1-f_{w_{|W_x|}})
         \\

        \vdots&\vdots  &\vdots  &\vdots  &\vdots  &\vdots  &\vdots  &\vdots \\

         -q_{p_{L_0},1}f_{1}&-q_{p_{L_0},2}f_{2} &\cdots  &-q_{p_{L_0},N_{y}}f_{N_{y}}  &-q_{p_{L_0},w_1}(1-f_{w_1})  &-q_{p_{L_0},w_2}(1-f_{w_2})  &\cdots  &-q_{p_{L_0},w_{|W_x|}}(1-f_{w_{|W_x|}})
         \\

            \hline
    
        q_{1,1}\beta_{-,1}&q_{1,2}\beta_{-,2}  &\cdots  &q_{1,N_{y}}\beta_{-,N_{y}}  &0  &0  &\cdots  &0 \\
    
        q_{2,1}\beta_{-,1}&q_{2,2}\beta_{-,2}  &\cdots  &q_{2,N_{y}}\beta_{-,N_{y}}  &0  &0  &\cdots  &0 \\
    
        \vdots&\vdots  &\vdots  &\vdots  &\vdots  &\vdots  &\vdots  &\vdots \\
    
         q_{N_{y},1}\beta_{-,1}&q_{N_{y},2}\beta_{-,2}  &\cdots  &q_{N_{y},N_{y}}\beta_{-,N_{y}}  &0 &0  &\cdots  &0 \\
     
      \end{array}\right].
      \nonumber\\
    \end{eqnarray}
    For simplification, we label the blocks of the matrix according to the division of the horizontal and vertical dividing lines as
    \begin{eqnarray}
        \widetilde{M}_R^3=\begin{bmatrix}
            X&Q_{\rm sub}\Lambda_{(1-f_w)}\\
            Q\Lambda_{\beta_-}&0
        \end{bmatrix},\label{eq:Qsub}
    \end{eqnarray}
    where $\Lambda_{\beta_-}={\rm diag}(\beta_{-,1},\beta_{-,1},\cdots,\beta_{-,N_{y}})$ and $\Lambda_{(1-f_w)}={\rm diag}[(1-f_{w_1}),(1-f_{w_2}),\cdots,(1-f_{w_{|W_x|}})]$. 
    In our consideration, the ranks of these submatricies satisfies
    $${\rm rank}(Q\Lambda_{\beta_-})={\rm rank}(Q)=N_{y}$$
    as $\beta_{\pm,y'}$ need to be finite to give valid solutions of eigenstates; 
    and 
    $${\rm rank}(Q_{\rm sub}\Lambda_{(1-f_w)})={\rm rank}(Q_{\rm sub}),$$
    as having any $1-f_w=0$ (an diagonal element of $\Lambda_{(1-f_w)}$) means that $\beta_{-,w}=\beta_{+,w}$, which uniquely determines their values and may give only a finite number of eigenstates that do not represent possible skin states.
    Therefore, the rank of $\widetilde{M}_R^3$ is restricted by
    \begin{eqnarray}
       N_{y}+{\rm rank}(Q_{\rm sub})= {\rm rank}(Q\Lambda_{\beta_-})+{\rm rank}(Q_{\rm sub}\Lambda_{(1-f_w)})=
        {\rm rank}(\begin{bmatrix}0&Q_{\rm sub}\Lambda_{(1-f_w)}\\
                    Q\Lambda_{\beta_-}&0
        \end{bmatrix})\nonumber\\
            \leq {\rm rank}(\widetilde{M}_R^3) 
            \leq{\rm rank}
            (\begin{bmatrix}
                X\\Q\Lambda_{\beta_-}
           \end{bmatrix})+
           {\rm rank}
           (\begin{bmatrix}
                 Q_{\rm sub}\Lambda_{(1-f_w)}\\0
            \end{bmatrix})=N_{y}+{\rm rank}(Q_{\rm sub}),
    \end{eqnarray}
    therefore 
    \begin{eqnarray}
    \label{eq:rankeqt}
        {\rm rank}(\widetilde{M}_R)=N_{y}+{\rm rank}(Q_{\rm sub}).
    \end{eqnarray}
    As a result, the condition to have skin states, i.e.,  ${\rm rank}(\widetilde{M}_R)<N_{y}+|W_x|$, is reduced to
    \begin{eqnarray}
    {\rm rank}(Q_{\rm sub})<|W_x|.\label{eq:rank_Q_sub}
    \end{eqnarray}    
    Since $Q_{\rm sub}$ has $|W_x|$ columns and $L_0$ rows, 
    Eq.~\eqref{eq:rank_Q_sub} is satisfied when $|W_x|>L_0$; 
    or, when 
    \begin{eqnarray}
    {\rm det}[Q_{\rm minor}]=0\label{eq:zero_minor}
    \end{eqnarray}
    with $Q_{\rm minor}$ a $|W_x|\times |W_x|$ square submatrix of $Q_{\rm sub}$, if $|W_x|\leqslant L_0$. 
    We emphasize that Eq.~\eqref{eq:zero_minor} is sufficient, but not necessary for Eq.~\eqref{eq:rank_Q_sub} to hold.
    However, since $Q_{\rm minor}$ is a minor of $Q$, we may first identify zero minors of $Q$, then search for $Q_{\rm sub}$ that contains such a zero minor.
   Note that $Q_{\rm sub}$ is a $L_0\times |W_x|$ submatrix of $Q$ composed by elements in its $(w_1, w_2, \cdots, w_{|W_x|})$th columns and $(p_1, p_2, \cdots, p_{L_0})$th rows. In other words, once the parameters of the model are fixed (so that $Q$ is fixed),
     $Q_{\rm sub}$ is uniquely determined by the spectral winding number of the given energy $E$ and the configuration of defects (which determine the values of $b_r$).
     Thus, the correspondence between zero minors of $Q$, MBCs, spectral winding numbers,  and the emergence of skin defect states is established as below: 
     \\
     
     \textbullet \quad Consider a system whose $Q$ matrix has at least one zero minor, ${\rm det}[Q_{\rm minor}]=0$. 
     A submatrix of $Q$, $Q_{\rm sub}$, is uniquely determined by the configuration of defects and the spectral winding number $W_x$ [see Eqs.~\eqref{eq:MR1} to \eqref{eq:Qsub}].
     Skin defect states emerge within the region characterized by $W_x$, provided $Q_{\rm minor}$ is a submatrix of $Q_{\rm sub}$.
     \\

    \begin{figure}
        \includegraphics[width=1\textwidth]{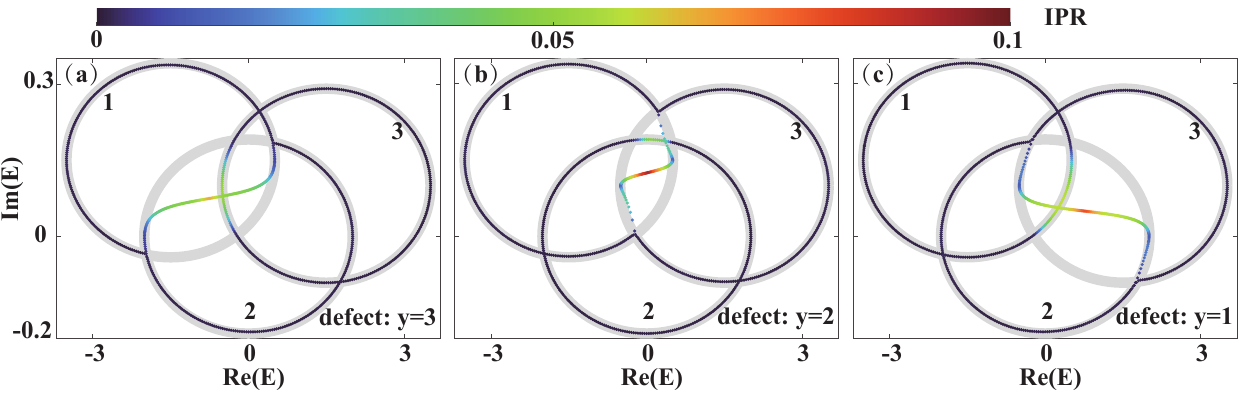}  
        \caption{Verification of the generalized correspondence with a lattice with $t_R=1.1$, $t_L=1/1.1$, $N_y=3$, and $N_{x}=300$, and the coupling matrix $J$ is given by Eq.\eqref{eq:gBBC_param1}. Colors indicate the inverse participation ration ${\rm IPR}=\sum_{x,y}|\psi_{x,y}|^4$ of each eigenstate in the presence of defects. Gray loops display the spectra under PBCs (without defects). The three ellipses of the PBC spectra are labeled by numbers 1, 2, and 3. A single defect is placed at (a) $y=3$, (b) $y=2$, and (c) $y=1$. 
        Note that the location of defects are defined for the original spatial coordinate of $\psi_{x,y}$, and different ellipses of the PBC spectrum correspond to the recombined components of $\phi_{x,y'}$. 
        That is, the locations of defects does not necessarily need to be the same as the numbers of the ellipses.
        }
        \label{fig:gBBC}
    \end{figure}

    In Fig. \ref{fig:gBBC}, we provide some concrete examples with $N_y=3$ to verified our analysis, where we set the coupling matrix $J=Q\Lambda Q^{-1}$ as:
    \begin{eqnarray}
        \label{eq:gBBC_param1}
        J=\begin{bmatrix}
            \frac{3}{4} + \frac{1}{20}i  & -\frac{3}{2} & -\frac{3}{4} + \frac{1}{20}i\\
            - \frac{3}{4} - \frac{1}{20}i& \frac{1}{10}i& -\frac{3}{4} + \frac{1}{20}i\\
            \frac{3}{4} + \frac{1}{20}i  & -\frac{3}{2} & -\frac{3}{4} + \frac{1}{20}i
        \end{bmatrix},~
        Q=\begin{bmatrix}
            \frac{1}{\sqrt{3}} & \frac{1}{\sqrt{3}}& - \frac{1}{\sqrt{3}}\\
            \frac{1}{\sqrt{3}}& \frac{1}{\sqrt{3}}& \frac{1}{\sqrt{3}}\\
            \frac{1}{\sqrt{3}}&  -\frac{1}{\sqrt{3}}& - \frac{1}{\sqrt{3}}
        \end{bmatrix}.
    \end{eqnarray}
   It is obvious that the submatrixes $Q_{[1,2],[1,2]},Q_{[2,3],[2,3]},Q_{[1,3],[1,3]}$ are singular, corresponding to zero minors of $Q$, where $Q_{\mathbf{ r},\mathbf{ c}}$ denotes the submatrix formed by $\mathbf{r}$ rows and $\mathbf{c}$ columns. 
   According to previous analysis, ${\rm det}(Q_{[1,2],[1,2]})=0$ corresponds to the emergence of skin defect states in the region enclosed by the first and second PBC ellipses, when a single defect is placed at $y=3$, which is the case of Fig.\ref{fig:gBBC}.(a). 
   Similarly, $Q_{[2,3],[2,3]}$ and $Q_{[1,3],[1,3]}$ correspond to (b) and (c), respectively.

\bigskip

    Finally, to complete this section, we move back to the case where any minor of $Q$ is nonzero, which yields ${\rm rank}(Q_{\rm sub})={\rm min}(L_0, |W_x|)$, the smaller value between the numbers of its columns and rows.
    According to Eq.~\eqref{eq:rankeqt}, we have ${\rm rank}(\widetilde{M}_R)=N_{y}+ L_0$ when $|W_x|>L_0$,
    thus satisfying Eq~\eqref{eq:ineq_1} and suggesting the existence of skin defect states localized toward $x=N_x$ in this case.
    However, when $|W_x|\leq L_0$,
    the rank of $\widetilde{M}_R$ is given by ${\rm rank}(\widetilde{M}_R)=N_{y}+ |W_x|$,
    which violates Eq.~\eqref{eq:ineq_1} and thus forbids the emergence of skin defect states.



    \subsection{V. Skin defect states and their topological origin in a generalized model}\label{sec:generalized}
    \subsubsection{A. generalized 1D models}
In the main text and previous discussion in this supplemental file, we have mapped our two-dimensional (2D) model in Eq.~(1) in the main text to a 1D Hamiltonian
\begin{eqnarray}
    \label{eq:sub_Hamiltonian}
    H\rightarrow H_{\rm 1D}&=&\sum_{x=1}^{N_{\rm x}-1}(t_R\hat{C}^{\dagger}_{x+1}\hat{C}_{x}+t_L\hat{C}^{\dagger}_{x}\hat{C}_{x+1})
    \\
    &+&(t_R\hat{C}^{\dagger}_{1}B\hat{C}_{N_{\rm x}}+t_L\hat{C}^{\dagger}_{N_{x}}B\hat{C}_{1})
    +\sum_{x=1}^{N_{\rm x}}(\hat{C}^{\dagger}_xJ\hat{C}_x)
    \nonumber
\end{eqnarray}
with $\hat{C}_x=\left[\hat{c}_{x,1}~\hat{c}_{x,2}~\cdots~c_{x,y}~\cdots~c_{x,N_y}\right]^{\rm T}$,
and different $y$ lattices are taken as different components within unit cells labeled by $x$.
Since the 2D model possesses translational symmetry along $y$ direction (apart from the defects), the nearest-neighbor hopping in $H_{\rm 1D}$ is the same for different $y$ lattices, described by two scalars $t_R$ and $t_L$.
In this section, we extend our analysis to general 1D non-Hermitian models
described by the Hamiltonian
 \begin{eqnarray}
 \label{eq:H_g}
 H_g&=&\sum_{x=1}^{N_{x}-1}(C^{\dagger}_{x+1}J_R C_{x}+C^{\dagger}_{x}J_LC_{x+1})
    \\
    &+&(C^{\dagger}_{1}BJ_RC_{N_{x}}+C^{\dagger}_{N_{x}}BJ_LC_{1})
    +\sum_{x=1}^{N_{x}}(C^{\dagger}_xJC_x),
    \nonumber
\end{eqnarray}
where the translational symmetry along $y$ direction is broken and the nearest-neighbor hopping $t_{R}$ and $t_{L}$ are replaced by
matrices $J_{R}$ and $J_{L}$, representing multiple hopping between $(x,y_1)$ and $(x\pm1,y_{2,\pm})$ without any restriction of $y_1$ and $y_{2,\pm}$.
This model involves arbitrarily long-range hopping along $y$ direction, but only nearest-neighbor hopping along $x$ direction.
Nevertheless,
long-range hopping along $x$ direction can be mapped to nearest-neighbor ones by considering larger unit cells, thus $H_g$ already includes all possible 1D lattices with bulk-translational symmetry along $x$ direction.
We will further discussion how arbitrary higher-dimensional systems can be mapped to this 1D general Hamiltonian later in Sec. VI.

\subsubsection{B. A demonstrative example}
\begin{figure}
    \includegraphics[width=0.4\linewidth]{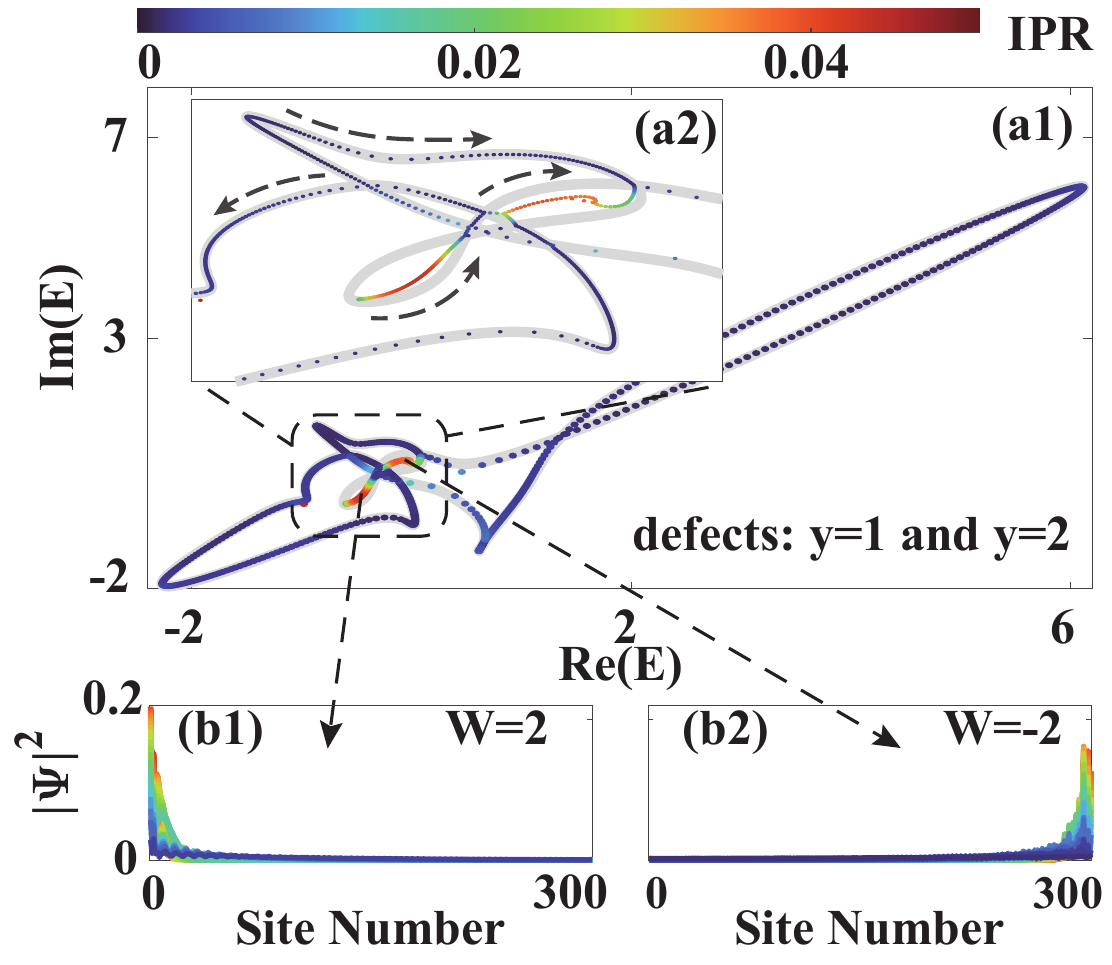}  
    \caption{(a1) and (a2) complex spectra and its local magnification of the generalized Hamiltonian $H_g$ with $N_y=3$ and $N_x=300$, and two defects at $y=1$ and $y=2$. Colors indicate the inverse participation ratio ${\rm IPR}=\sum_{x,y}|\psi_{x,y}|^4$ of each eigenstate, gray loops display the PBC spectrum in the absence of defects. (b1) and (b2) the distribution of eigenstates within the regions with $W=-2$ and $W=2$, respectively, with colors indicate the IPR of states.
    Other parameters are listed in Eq.~\eqref{eq:param_set1}.
    }
    \label{fig:general}
\end{figure}
Before moving on to deducing the correspondence between spectral winding topology and skin defect states for $H_g$,
we first demonstrate in Fig. \ref{fig:general} numerical results of this Hamiltonian with $N_y=3$ and a set of randomly chosen hopping parameters,
\begin{eqnarray}
   J_R=\begin{bmatrix}
     0.934+0.265i & 0.726+0.253i & 0.722+0.453i\\
     0.957+0.667i & 0.259+1.043i & 1.209+0.406i\\
     0.627+0.634i & 0.355+0.414i & 0.524+1.140i
    \end{bmatrix},\nonumber\\    J=\begin{bmatrix}
        0.199+0.628i & 1.043+1.016i & 0.625+0.092i\\
        0.600+0.579i & 0.816+0.675i & 0.715+0.637i\\
        0.896+0.576i & 0.427+0.776i & 0.510+0.938i
    \end{bmatrix},\nonumber\\
    J_L=\begin{bmatrix}
        0.487+0.827i & 0.673+0.532i & 0.653+0.107i\\
        0.528+0.529i & 0.454+1.201i & 0.472+0.697i\\
        0.964+1.049i & 1.074+1.034i & 0.845+0.927i
        \label{eq:param_set3}
    \end{bmatrix}.
\end{eqnarray}
It is seen that the full-PBC spectrum simultaneously supports both positive and negative spectral winding numbers at different reference energies, but no longer forms a series of ellipses,
making it more difficult to analyze
the relation between spectral winding number and skin defect states.
Nevertheless, we observe that skin defect states localized at the left (right) edge emerge only in the region with $W>L_0$ ($-W>L_0$).


\subsubsection{C. Deduction of the topological origin of skin defect states}
Starting from the Hamiltonian of Eq.~\eqref{eq:H_g},The eigenequation $H|\Psi\rangle=E|\Psi\rangle$, 
with $|\Psi\rangle=\sum_{n=1}^{N_{y}}\sum_{x=1}^{N_{x}}\psi_{n,x}c^\dagger_{n,x}|0\rangle$ and $|0\rangle$ the vaccum state, can be expressed as the bulk equations
\begin{eqnarray}
    \label{eq:bulk_sup}
    E\Psi_x=J\Psi_x+J_R\Psi_{x-1}+J_L\Psi_{x+1}
\end{eqnarray}
with $x=2,3,\cdots,N_{x}$, and the boundary equations
\begin{eqnarray}
    \label{eq:boundary_0_sup}
    E\Psi_1&=&J\Psi_1+BJ_R\Psi_{N_{x}}+J_L\Psi_2,\\
    \label{eq:boundary_Nc+1_sup}
    E\Psi_{N_{x}}&=&J\Psi_{N_{x}}+J_R\Psi_{N_{x}-1}+BJ_L\Psi_{1}.
\end{eqnarray}
We may also extend the bulk equations to the boundary as
\begin{eqnarray}
    \label{eq:extended_bulk_1_g_sup}
    E\Psi_1&=&J\Psi_1+J_R\Psi_0+J_L\Psi_2,\\
    \label{eq:extended_bulk_2_g_sup}
    E\Psi_{N_{x}}&=&J\Psi_{N_{x}}+J_R\Psi_{N_{x}-1}+J_L\Psi_{N_{x}+1}.
\end{eqnarray}
Compare Eq.\eqref{eq:boundary_0_sup}, Eq.\eqref{eq:boundary_Nc+1_sup} and Eq.\eqref{eq:extended_bulk_1_g_sup}, Eq.\eqref{eq:extended_bulk_2_g_sup}, we have the following boundary conditions:
\begin{eqnarray}
    \label{eq:boundary_1_g_sup}
    J_R\Psi_0&=&BJ_R\Psi_{N_{x}},\\
    \label{eq:boundary_2_g_sup}
    J_L\Psi_{N_{x}+1}&=&BJ_L\Psi_{1}.
\end{eqnarray}

We first deal with the bulk equations. Substituting the ansatz $\Psi_x=\beta^x\Psi_0$ in to Eq.\eqref{eq:bulk_sup}, we have
\begin{eqnarray}
    \label{eq:bulk_beta_supmat}
    (J_R\beta^{-1}+(J-EI)+J_L\beta)\Psi_0=0,
\end{eqnarray}
where $I$ is the $N_{y}\times N_{y}$ identity matrix. The condition to have nontrivial solutions of $\Psi_0$ is given by
\begin{eqnarray}
    \label{eq:det_general_supmat}
    {\rm det}(J_R\beta^{-1}+(J-EI)+J_L\beta)=0,
\end{eqnarray}
which can be viewed as a $2N_{y}$-degree equation of $\beta$.
We label the solutions of $\beta$ as $\beta_{r}$ with ${r}=1,2,...,2N_{y}$, each as a function of the energy $E$.
On the other hand, by taking $E$ as a variable and letting $\beta=e^{ik_x}$, $k\in[0,2\pi)$, Eq.~\eqref{eq:det_general_supmat} becomes
\begin{eqnarray}
    \label{eq:det_g}
    {\rm det}(J_Re^{-ik_x}+(J-EI)+J_Le^{ik_x})=0,
\end{eqnarray}
which gives the PBC spectrum in the thermodynamic limit.
In other words,
\\

\textbullet \quad $|\beta|=1$ may be satisfied only when $E$ falls on the PBC spectrum.
\\
\\
Therefore, denoting 
the numbers of $\beta$s satisfying $|\beta|<1$ and $|\beta|>1$ as $N_{+}$ and $N_{-}$, respectively,
we always have $N_{+}+N_{-}=2N_{y}$ as long as $E$ does not fall on the full-PBC spectrum.

Next we shall establish the exact BBC for the general model in two steps:
\\

\textbullet \quad the correspondences between the spectral winding number $W_x$ and $N_\pm$ is given by $N_{\pm}=N_{y}\pm W_x$; 
\\

\textbullet \quad skin modes emerge only when $N_+>{\rm rank}(\widetilde{M}_+)$ or $N_->{\rm rank}(M_-)$, 
where $\widetilde{M}_+$ and $M_{-}$ two auxiliary matrices 
with their ranks no larger than $N_{y}+L_0$ (the number of rows with nonzero elements).
\\
\\
Combining these results, we obtain the exact BBC for the general model, 
where left-localized and right-localized skin modes emerge for eigenenergies in the regions with $W> L_0$ and $-W>L_0$, respectively.

\subsubsection{D. Relation between $N_{\pm}$ and the spectral winding number $W_x$}
Firstly, we note that the left side of Eq. \eqref{eq:det_general_supmat} can be written as:
\begin{eqnarray}
    \label{eq:poly_sup}
    {\rm det}(J_R\beta^{-1}+(J-EI)+J_L\beta)=\frac{P_{2N_{y},E}(\beta)}{\beta^{N_{y}}},
\end{eqnarray}
where $P_{2N_{y}}(\beta,E)$ is a $2N_{y}$-th order polynomial of $\beta$, with $E$ taken as a parameter. 
The relation of $N_{\pm}=N_{y}\pm W_x$ can be derived from Cauchy principle, similar to the derivation in Ref.~\cite{zhang2020correspondence}. 
According to Cauchy principle, 
the spectral winding number for a given energy $E$ satisfies
$W_x(E)=N_{\rm zeros}-N_{\rm poles}$,
where $N_{\rm zeros}$ and $N_{\rm poles}$ are the
the counting of zeros and poles of $P_{2N_{y}}(\beta,E)$ enclosed by the unit circuit in the complex space of $\beta$ (i.e., the Brillouin zone of $\beta=e^{ik_x}$), weighted by respective orders.
Note that $N_{\rm poles}=N_{y}$ following Eq.~\eqref{eq:poly_sup}, 
and $N_{\rm zeros}=N_{+}$ since the zeros enclosed by BZ correspond to $|\beta|<1$.
Hence we obtain the condition $N_{\pm}=N_{y}\pm W_x$.

In Fig.~\ref{fig:phase_d}, we numerically verify this relation
by solving Eq.\eqref{eq:det_general_supmat}
for $\beta$s, with the coupling matrix $J_R,J,J_L$ stochastically generated as

(a):
\begin{eqnarray}
    J_R=\begin{bmatrix}
        0.89+0.11i & 0.82+0.53i\\
        0.33+0.60i & 0.04+0.42i
    \end{bmatrix}
    \quad
    J=\begin{bmatrix}
        0.34+0.52i &  0.44+0.65i\\
        0.62+0.58i &  0.74+0.99i
    \end{bmatrix}
    \quad
    J_L=\begin{bmatrix}
        0.82+0.05i & 0.88+0.80i\\
        0.41+0.72i & 0.82+0.74i
    \end{bmatrix},
    \label{eq:param_set1}
\end{eqnarray}

(b):
\begin{eqnarray}
    J_R\begin{bmatrix}
        0.10 + 0.53i & 0.27 + 0.48i\\
        0.79 + 0.64i & 0.61 + 0.60i
     
    \end{bmatrix}
    \quad
    J=\begin{bmatrix}
        0.29 + 0.76i &  0.48 + 0.41i\\
        0.38 + 0.63i &  0.67 + 0.80i
    \end{bmatrix}
    \quad
    J_L=\begin{bmatrix}
        0.26 + 0.01i &  0.72 + 0.73i\\
        0.12 + 0.77i &  0.23 + 0.13i
     \label{eq:param_set2}
    \end{bmatrix}
\end{eqnarray}

(c):
the parameters in Eq.~\eqref{eq:param_set1}.
\\

It is obviously seen in Fig.\ref{fig:phase_d}, $N_{-}=N_{y}-W_x$ and therefore $N_{+}=N_{y}+W_x$, verifying the above deduction.
\begin{figure}
    \includegraphics[width=1\textwidth]{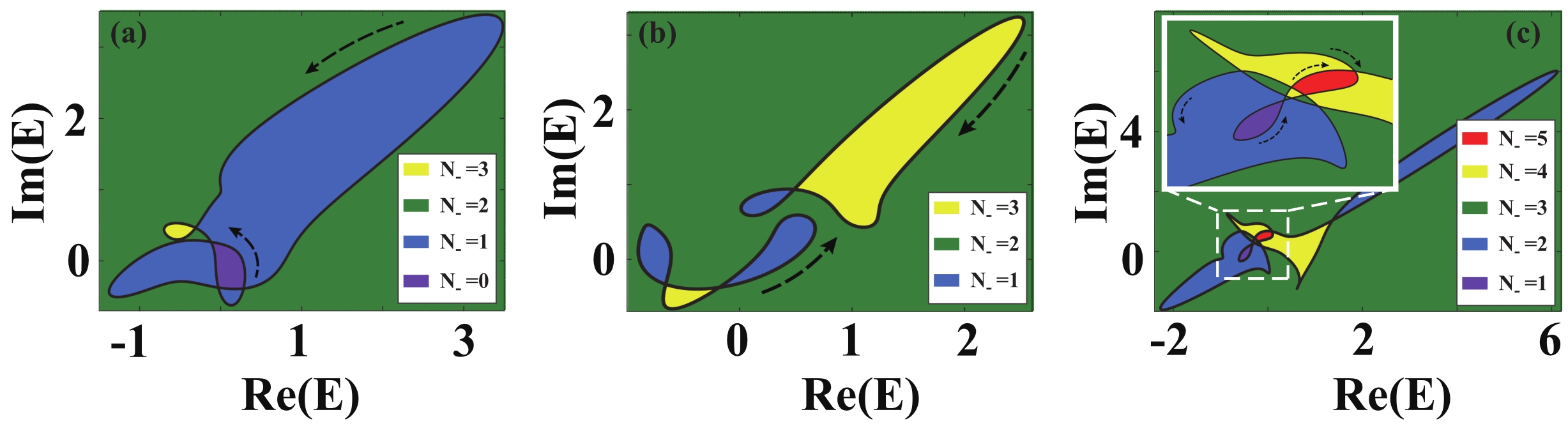}  
    \caption{Examples of the spectral winding number and $N_-$ for different energy $E$.
    The black loops show the PBC spectra, with the parameters given by (a) Eq.\eqref{eq:param_set1}, (b) Eq.\eqref{eq:param_set2}, and (c) Eq.\eqref{eq:param_set3}. 
    The size of the lattice is chosen to be $N_y=2$ in (a) and (b), and $N_y=3$ in (c).
    Dash arrows depict the winding directions of the PBC spectra, and colors depict the values of $N_-$.
    }
    \label{fig:phase_d}
\end{figure}

\subsubsection{E. Relation between $N_\pm$ and the emergence of skin defect states}
Similarly to the analysis for the previous simple model, 
the relation between $N_\pm$ and the emergence of skin defect stat4es can be obtained by
considering the boundary equations \eqref{eq:boundary_1_g_sup} and \eqref{eq:boundary_2_g_sup}. 
The formal solution of eigenstates for the general model shall be given by linear combinations of different solutions of $\beta$s,
\begin{eqnarray}
\psi_{x,y}=\sum_m^{2N_{y}}\gamma_{m,y}\beta_m^x.
\end{eqnarray}
To apply a similar analysis as for the previous model, we rewrite the the coefficient $\gamma_{m,y}$ as 
$\gamma_{m,y}=\gamma_m v_{m,y}$,
and the total wavefunction as
\begin{eqnarray}
    \label{eq:expand_psix_general_supmat}
    \Psi_{x}=\sum_{m=1}^{2N_{y}}\gamma_m\beta_m^x\mathbf{V}_m,
\end{eqnarray}
with $\mathbf{V}_m=[v_{m,1} v_{m,2},... v_{m,N_{y}}]^{\rm T}$ a column vector.
Similar to the treatment of Eq.~(9) in the main text, we rewrite $\Psi_x$ as
\begin{eqnarray}
    \label{eq:Tform_psix_g_supmat}
    \Psi_x=T\Lambda_\beta^x\Gamma,
\end{eqnarray}
where
\begin{eqnarray}
    \label{eq:T_g_supmat}
    T&=&\begin{bmatrix}
        \mathbf{V}_1&\mathbf{V}_2&\cdots&\mathbf{V}_{2N_{y}}
    \end{bmatrix},\\
    \label{eq:gamma_g_supmat}
    \Gamma&=&\begin{bmatrix}
        \gamma_1&\gamma_2&\cdots&\gamma_{2N_y}
    \end{bmatrix},\\
    \label{eq:Lambda_g_supmat}
    \Lambda_\beta&=&{\rm diag}(\beta_1,\beta_2,\cdots,\beta_{2N_{y}}),
\end{eqnarray}
therefore, the two boundary equations Eq.\eqref{eq:boundary_1_g_sup} and Eq.\eqref{eq:boundary_2_g_sup} can be written as:
\begin{eqnarray}
    \label{eq:boundary_Tform_g_supmat}
    \begin{bmatrix}
        J_R\Psi_0-BJ_R\Psi_{N_{x}}\\
        J_L\Psi_{N_{x}+1}-BJ_L\Psi_{1}
     \end{bmatrix}=
      \begin{bmatrix}
        J_RT-BJ_RT\Lambda_{\beta}^{N_{x}}\\
        J_LT\Lambda_{\beta}^{N_{x}+1}-BJ_LT\Lambda_{\beta}
     \end{bmatrix}\Gamma=
     M\Gamma=0,
\end{eqnarray}
we label the elements of $J_RT$ as $\widetilde{j^R}_{r,m}$ and the elements of $J_LT$ as $\widetilde{j^L}_{r,m}$ with $r=1,2,\cdots,N_{y}$ and $m=1,2,\cdots,2N_{y}$.
The elements can be simplified in the thermodynamic limit ($N_{x}\rightarrow \infty$) as
\begin{eqnarray}
    \label{eq:M_elements1_simplify_general_supmat}
    M_{r,m}&=&\widetilde{j^R}_{r,m}-\widetilde{j^R}_{r,m}\beta_{m}^{N_{x}}b_r\\
    &\underset{N_{x}\rightarrow \infty}{=}&\begin{cases}
        -\widetilde{j^R}_{r,m}\beta^{N_{x}}_{m}b_r,\quad {\rm when}~|\beta_{m}|>1;\nonumber\\
        \widetilde{j^R}_{r,m},\qquad\qquad\ \, {\rm when}~|\beta_{m}|<1;
       \end{cases}\\
    \label{eq:M_elements2_simplify_general_supmat}
    M_{r+N_{x},m}&=&\widetilde{j^L}_{r,m}\beta_{m}^{N_{x}+1}-\widetilde{j^L}_{r,m}\beta_{m}b_r\\
    &\underset{N_{x}\rightarrow \infty}{=}&\begin{cases}
        \widetilde{j^L}_{r,m}\beta_{m}^{N_{x}+1},\quad {\rm when}~|\beta_{m}|>1;\nonumber\\
       -\widetilde{j^L}_{r,m}\beta_{m}b_r,\quad {\rm when}~|\beta_{m}|<1.
       \end{cases}
\end{eqnarray}
Assuming $|\beta_{m}|>1$ for $m=m_{-,1},m_{-,2},\cdots,m_{-,N_-}$, and $|\beta_{m}|<1$ for $m=m_{+,1},m_{+,2},\cdots,m_{+,N_+}$,
the submatrix $M_-$ and $M_+$, corresponding to $|\beta_m|>1$ and $|\beta_m|<1$ respectively, can be expanded as
\begin{eqnarray}
    \label{eq:M_-_g_supmat}
    \widetilde{M}_-=M_-\Lambda_{\beta_-}^{-N_{x}}=
  \left[\begin{array}{cccc}
    -\widetilde{j^R}_{1,m_{-,1}}b_{1}&-\widetilde{j^R}_{1,m_{-,2}}b_{1}  &\cdots  &-\widetilde{j^R}_{1,m_{-,N_-}}b_{1}   \\

    -\widetilde{j^R}_{2,m_{-,1}}b_{2}&-\widetilde{j^R}_{2,m_{-,2}}b_{2}  &\cdots  &-\widetilde{j^R}_{2,m_{-,N_-}}b_{2}   \\

    \vdots&\vdots  &\vdots  &\vdots   \\

    -\widetilde{j^R}_{N_{y},m_{-,1}}b_{N_{y}}&-\widetilde{j^R}_{N_{y},m_{-,2}}b_{N_{y}}  &\cdots  &-\widetilde{j^R}_{N_{y},m_{-,N_-}}b_{N_{y}}   \\

        \hline

  \widetilde{j^L}_{1,m_{-,1}}\beta_{m_{-,1}}&\widetilde{j^L}_{1,m_{-,2}}\beta_{m_{-,2}}  &\cdots  &-\widetilde{j^L}_{1,m_{-,N_-}}\beta_{m_{-,N_-}}   \\

    \widetilde{j^L}_{2,m_{-,1}}\beta_{m_{-,1}}&\widetilde{j^L}_{2,m_{-,2}}\beta_{m_{-,2}}  &\cdots  &\widetilde{j^L}_{2,m_{-,N_-}}\beta_{m_{-,N_-}}   \\

    \vdots&\vdots  &\vdots  &\vdots   \\

    \widetilde{j^L}_{N_{y},m_{-,1}}\beta_{m_{-,1}}&\widetilde{j^L}_{N_{y},m_{-,2}}\beta_{m_{-,2}}  &\cdots  &\widetilde{j^L}_{N_{y},m_{-,N_-}}\beta_{m_{-,N_-}}   \\
 
  \end{array}\right]_{2N_{y}\times(N_{y}-W)},
\end{eqnarray}
with $\Lambda_{\beta_-}={\rm diag}(\beta_{m_{-,1}},\beta_{m_{-,2}},\cdots,\beta_{m_{-,N_-}})$,
and
\begin{eqnarray}
    \label{eq:M_+_g_supmat}
    M_+=
  \left[\begin{array}{cccc}
    \widetilde{j^R}_{1,m_{+,1}}&\widetilde{j^R}_{1,m_{+,2}} &\cdots  &\widetilde{j^R}_{1,m_{+,N_+}}   \\

    \widetilde{j^R}_{2,m_{+,1}}&\widetilde{j^R}_{2,m_{+,2}}  &\cdots  &\widetilde{j^R}_{2,m_{+,N_+}}   \\

    \vdots&\vdots  &\vdots  &\vdots   \\

    \widetilde{j^R}_{N_{y},m_{+,1}}&\widetilde{j^R}_{N_{y},m_{+,2}}  &\cdots  &\widetilde{j^R}_{N_{y},m_{+,N_+}}   \\

        \hline

  -\widetilde{j^L}_{1,m_{+,1}}\beta_{m_{+,1}}b_{1}&-\widetilde{j^L}_{1,m_{+,2}}\beta_{m_{+,2}}b_{1}  &\cdots  &-\widetilde{j^L}_{1,m_{+,N_+}}\beta_{m_{+,N_+}}b_{1}   \\

    -\widetilde{j^L}_{2,m_{+,1}}\beta_{m_{+,1}}b_{2}&-\widetilde{j^L}_{2,m_{+,2}}\beta_{m_{+,2}}b_{2}  &\cdots  &-\widetilde{j^L}_{2,m_{+,N_+}}\beta_{m_{+,N_+}}b_{2}   \\

    \vdots&\vdots  &\vdots  &\vdots   \\

    -\widetilde{j^L}_{N_{y},m_{+,1}}\beta_{m_{+,1}}b_{N_{y}}&-\widetilde{j^L}_{N_{y},m_{+,2}}\beta_{m_{+,2}}b_{N_{y}}  &\cdots  &-\widetilde{j^L}_{N_{y},m_{+,N_+}}\beta_{m_{+,N_+}}b_{N_{y}}   \\
 
  \end{array}\right]_{2N_{y}\times(N_{y}+W)},
  \nonumber\\
\end{eqnarray}
Similar to the analysis of the previous simple model (see the main text and Sec. II of this supplemental file), 
$\widetilde{M}_-$ has at most $N_{y}+L_0$ rows with nonzero elements,
thus Eq.~\eqref{eq:boundary_Tform_g_supmat} supports an infinite number of skin defect states localized toward $x=N_x$
when $N_-=N_{y}-W_x>N_{y}+L_0$, i.e., $-W_x>L_0$.
Similarly, skin defect states localized toward $x=1$ correspond to $N_+=N_{y}+W_x>N_{y}+L_0$, i.e., $W_x>L_0$.

Note that as for the previous model, a strict correspondence between skin defect states and the spectral winding number further requires $\widetilde{M}_-=N_{y}+L_0$. 
Otherwise, when 
\begin{eqnarray}
\widetilde{M}_-<N_{y}+L_0,\label{eq:rank_Mminus}
\end{eqnarray}
skin defect states may arise even when $|W_x|<L_0$, similar to Fig.~\ref{fig:gBBC} for the previous model with ${\rm rank}(\widetilde{M}_R)< N_{y}+|W_x|$. 
While it is difficult to find an explicit condition for Eq.~\eqref{eq:rank_Mminus} to hold,
we may conjecture that it is rather sensitive (e.g., with zero minors of $Q$ for the previous model), 
and the topological correspondence of skin defect states may also be recovered by disorder even in such cases.

\subsubsection{F. Mapping a $d$-dimensional lattices to a one-dimensional lattices}
To generalize our analysis to lattices in arbitrary dimension, we need to map
the lattice Hamiltonian to the one-dimensional one of Eq.~\eqref{eq:H_g}.
Consider a $d$-dimensional non-Hermitian lattice with $N_s$ components (sublattices, spins, orbitials, etc.) with in each unit cell. 
For convenience, we label the dimensions as $x_1$, $x_2$, ... $x_{d}$, and map the system to a one-dimensional (1D) one along $x$ direction.
Assuming along each direction $\alpha$, the maximal hopping ranges toward the left (smaller values of $\alpha$) and the right (larger values of $\alpha$) are given by $l_\alpha$ and $r_\alpha$, respectively.
The most general Hamiltonian can be written as
\begin{eqnarray}
&&H_d=\sum_{\bm{r}}\sum_{\bm{\delta}} \sum_{s,s'=1}^{N_s}t_{\bm{\delta},s,s'}\hat{c}_{\bm{r}+\bm{\delta},s}^\dagger \hat{c}_{\bm{r},s'},\\
&&\sum_{\bm{r}}=\sum_{x_1=1}^{N_{x_1}}\sum_{x_2=1}^{N_{x_2}}\dots\sum_{x_d=1}^{N_{x_d}},~~~~
\sum_{\bm{\delta}}=\sum_{\delta_{x_{1}}=-l_{x_{1}}}^{r_{x_{1}}} 
\sum_{\delta_{x_{1}}=-l_{x_{2}}}^{r_{x_{2}}}\dots
\sum_{\delta_{x_{1}}=-l_{x_{d}}}^{r_{x_{d}}}\nonumber
\end{eqnarray}
with $s=1,2,...,N_s$ describing different components, $\bm{r}=(x_1,x_2,...x_{d})$ the position of a unit cell,
and $\bm{\delta}=(\delta_{x_1},\delta_{x_2},...,\delta_{x_d})$.
Note that the ``hopping" parameters  $t_{\delta_{\alpha},s,s'}$ describe onsite potentials when $\bm{\delta}=0$ and $s=s'$.
As an example, the minimal 2D model in the main text has $d=2$, $l_{x_1}=r_{x_1}=l_{x_2}=r_{x_2}=1$ with $x_1\equiv x$ and $x_2\equiv y$, $s=s'=1$, and $N_s=1$.
Therefore $t_{\bm{\delta},s,s'}$ takes nonzero values only when
\begin{eqnarray} 
t_{\delta_x,1,1}=\begin{cases} 
      t_L, & (\delta_{x_1},\delta_{x_2})=(-1,0) \\
      t_R, &(\delta_{x_1},\delta_{x_2})=(1,0) \\
      j_L, & (\delta_{x_1},\delta_{x_2})=(0,-1) \\
      j_R, &(\delta_{x_1},\delta_{x_2})=(0,1)
   \end{cases}.
\end{eqnarray}

To apply our derivation of the emergence of skin defect states,
we need to map $H_d$ to a 1D model along $x_1\equiv x$ direction, with only nearest-neighbor hoppings (regarding unit cells) and on-site potentials.
To do so, we define a supercell as the sum of $M_x={\rm max}[r_x,l_x]$ unit cells. Thus, we may write down an equivalent 1D Hamiltonian as 
\begin{eqnarray}
H_{\rm 1D}=\sum_{x=1}^{N_x/M_x} \hat{C}_{x+1}^\dagger J_{R}\hat{C}_{x}+\hat{C}_{x}^\dagger J_{L}\hat{C}_{x+1}+\sum_{x=1}^{N_x/M_x}\hat{C}_{x}^\dagger  J\hat{C}_{x},\label{eq:mapping_1D}
\end{eqnarray}
with $x'$ the position of supercells,
$\hat{C}_x=\left[\hat{c}_{x,1}~\hat{c}_{x,2}~\hat{c}_{x,u}~\cdots~c_{x,N_u}\right]^{\rm T}$ the annihilation operators arranged in a column,
$$N_u=M_xN_s\prod_{n=1}^{d-1}N_{y_n}$$ the number of components in a supercell, 
and $u=1,2,...N_u$ labeling different components.
These components in each supercell is given by lattice sites and components 
in $M_x$ consecutive $(d-1)$-dimensional slices of the original model along $x$ direction, namely,
$$u \mapsto (m_x,s,x_2,x_3,...x_{d})$$ with $m_x=1,2,...M_x$.
Note that for a finite system, $N_x$ needs to be chosen as an integer multiple of $M_x$, so that the system has an integer number of supercells.

For the the minimal 2D model in the main text, the hopping matrices are given by $(J)_{u,u'}=j_L\delta_{u+1,u'}+j_R\delta_{u,u'+1}$,
$(J_L)_{u,u'}=t_L\delta_{u,u'}$, $(J_R)_{u,u'}=t_R\delta_{u,u'}$,
with size of the supercell $N_u=N_y$ (since $M_x=N_s=1$ and $d=2$).
The 1D model of Eq.~(3) in the main text has lifted the above restriction of $J$, and we have analytically solved the emergence of skin defect states in this model in the main text and in Sections II to IV of this supplemental file.
Extension of our analysis to the most general case without any restriction of $J$, $J_R$, and $J_L$ can be found in Sec. V.

\subsection{VI. Universality of the topological characterization of skin defect states}
Beyond the scope of analytical support of our work, we herein present numerical results of two distinct extensions to unveil the universality of the topological characterization of skin defect states unveiled in the main text.
\subsubsection{A. A domain wall between Hermitian and non-Hermitian subsystems}
We consider a 2D domain-wall system with a Hermitian lattice connected to the non-Hermitian one,
described by the Hamiltonian (before introducing defects)
\begin{eqnarray}
\label{eq:Hamiltonian}
    \label{eq:H_nH_coupling}
    H&=&H_{\rm nH}+H_{\rm H}+H_{\rm coupling},\\
    \label{eq:H_nH}
    H_{\rm nH}&=&\sum_{x=1}^{N^{\rm nH}_{\rm x}}(t_R\hat{C}^{\dagger}_{x+1}\hat{C}_{x}+t_L\hat{C}^{\dagger}_{x}\hat{C}_{x+1}),\\
    \label{eq:H_H}
    H_{\rm H}&=&\sum_{x=N^{\rm nH}_{\rm x}+1}^{N^{\rm nH}_{\rm x}+N^{\rm H}_{\rm x}}t_{\rm H}\hat{C}^{\dagger}_{x+1}\hat{C}_{x}+{\rm H.C.},\\
    \label{eq:H_coupling}
    H_{\rm coupling}&=&\sum_{x=1}^{{N^{\rm nH}_{\rm x}+N^{\rm H}_{\rm x}}}\hat{C}^{\dagger}_xJ\hat{C}_x,
\end{eqnarray}
with the column vector of annihilation operators
$\hat{C}_x=\left[\hat{c}_{x,1}~\hat{c}_{x,2}~\cdots~c_{x,N_y}\right]^{\rm T}$, and $(J)_{y_1,y_2}\equiv j_{y_1,y_2}=j_L\delta_{y_1+1,y_2}+j_R\delta_{y_1,y_2+1}$ the same hopping amplitudes along $y$ direction as in the 2D example in the main text.
The x-lengths of the Hermitian and non-Hermitian parts are denoted as $N^{\rm H}$ and $N^{\rm nH}$, respectively.
For simplicity, we have defined that ${N^{\rm nH}_{\rm x}+N^{\rm H}_{\rm x}}+1\equiv1$ for $x$ and $N_{\rm y}+1\equiv1$ for $y_1$ and $y_2$, which leads to the PBCs along both directions ($x$ and $y$). 

As translational symmetry is broken along x-axis by the domain wall, the spectral winding number in Eq. (8) in the main text becomes ill-defined.
Alternatively, we introduce an auxiliary Hamiltonian 
\begin{eqnarray}
H^{\rm au}&=&H_{\rm nH}^{\rm au}+H_{\rm coupling},\\
H_{\rm nH}^{\rm au}=&=&\sum_{x=1}^{N^{\rm nH}_{\rm x}+N^{\rm H}_{\rm x}}(t_R^{\rm au}\hat{C}^{\dagger}_{x+1}\hat{C}_{x}+t_L^{\rm au}\hat{C}^{\dagger}_{x}\hat{C}_{x+1}),
\end{eqnarray}
with the same ``cumulative strength" of non-reciprocity along $x$, i.e., $(t^{\rm au}_R/t^{\rm au}_L)^{N^{\rm nH}_{\rm x}+N^{\rm H}_{\rm x}}=(t_R/t_L)^{N^{\rm nH}_{\rm x}}$. The PBC spectrum of $H^{\rm au}$ is found to overlap with that of $H$ when the parameters further satisfy $t_Rt_L=t_R^{\rm au}t_L^{\rm au}=t_{\rm H}^2$. 
Thus, we can obtain an auxiliary winding number $\overline{W}_x^{\rm au}$ for $H^{\rm au}$, which also describes the spectral winding property of $H$. 

For simplicity, we choose $N_x^{\rm nH}=N_x^{\rm H}$, so that 
\begin{eqnarray}
t^{\rm au}_R=\sqrt{t_{\rm H}t_R},~
t^{\rm au}_L=\sqrt{t_{\rm H}t_L}.
\end{eqnarray}
In Fig. \ref{fig:H_NH} (a1) to (c1), we demonstrate eigenenergies marked by different colors according to the effective dimension $D_{\rm state}$ of their corresponding eigenstates, with black loops mark the critical value of $|\overline{W}_x^{\rm au}|=L_0/N_y$.
It is seen that eigenstates with eigenenergies inside the black loops exhibit stronger localization than the rest (indicated by their smaller $D_{\rm state}$).
Specifically, when the defects are placed within the non-Hermitian subsystem, 
eigenstates inside the loops are localized at the defects, manifesting skin defect states; while those outside the loops are localized at the domain wall, as shown in Fig. \ref{fig:H_NH}(a2). In particular, the skin defect states are found to exhibit stronger localization than the skin states at the domain wall (with $D_{\rm state}\approx 1.2\sim1.3$), as the defects introduce inhomogeneity and lead to localization also along $y$ direction.

In Fig. \ref{fig:H_NH}(b1) and (b2), we place the defects exactly at the domain wall. Eigenstates with the strongest localization (red) are also seen to mainly located inside the black loop, while localization of the some eigenstates outside loop is also strengthened (light blue). 
Finally, no skin defect state is observed when the defects are placed in the Hermitian subsystem, as shown in Fig. \ref{fig:H_NH}(c1) and (c2). However, normal skin states at the domain wall are found to exhibit stronger localization when their eigenenergies have $|\overline{W}_x^{\rm au}|>L_0/N_y$. In other words, the localization of skin states at the domain wall is enhanced by distant defects.

\begin{figure}
    \includegraphics[width=0.65\textwidth]{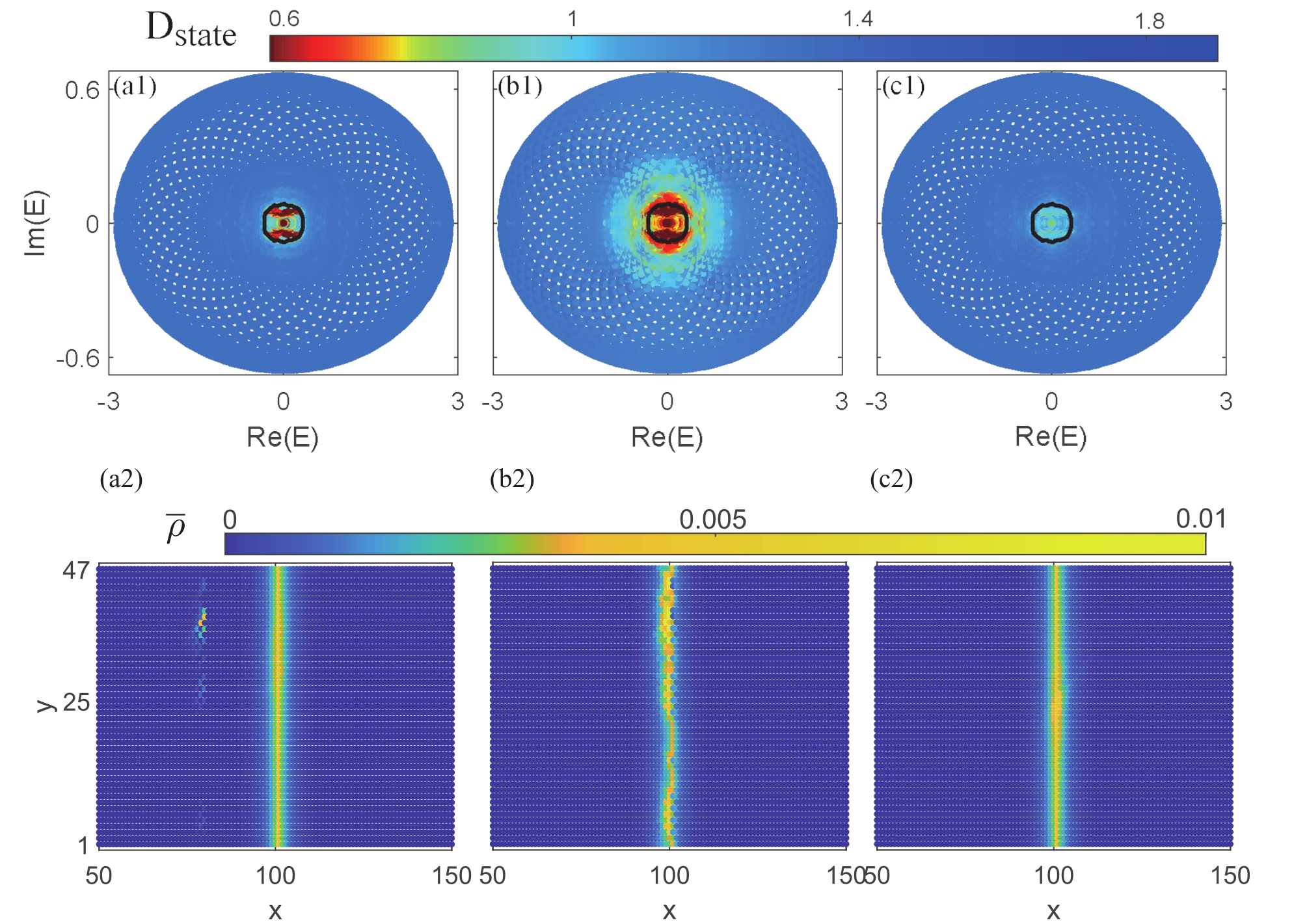}  
    \caption{Skin defect states and normal skin states at the domain wall of the model described by Eq.~\eqref{eq:H_nH_coupling}. 
    (a1) to (c1) Spectra of the model with defects at $x=80$, $x=100$, and $x=120$, respectively.
    Other parameters are
    $t_R=1.2$, $t_L=0.5$, $t=\sqrt{0.6}$, $j_R=0.6$, $j_L=1$, $N_{\rm nH}=N_{\rm H}=100$, $N_y=47$, and  $L_d=16$ defective sites along $y$ direction, located at $y=3,5,7,9,15,23,25,27,29,32,35,37,39,40,44,45$.
    The eigenenergies are colored according to $D_{\rm state}$, the effective dimension of corresponding eigenstates. 
    Black loops separate regions with the auxiliary winding number $|\overline{W}_x^{\rm au}|$ larger and smaller than $L_0/N_y$ (with $L_0=N_y-L_d$).
    (a2) to (c2) average distribution of all eigenstates, $\overline{\rho}=\sum_n|\psi_{x,y}^n|^2/N_y$, corresponding to (a1) to (c1), respectively.
    }
    \label{fig:H_NH}
\end{figure}
\begin{figure}
    \includegraphics[width=0.65\textwidth]{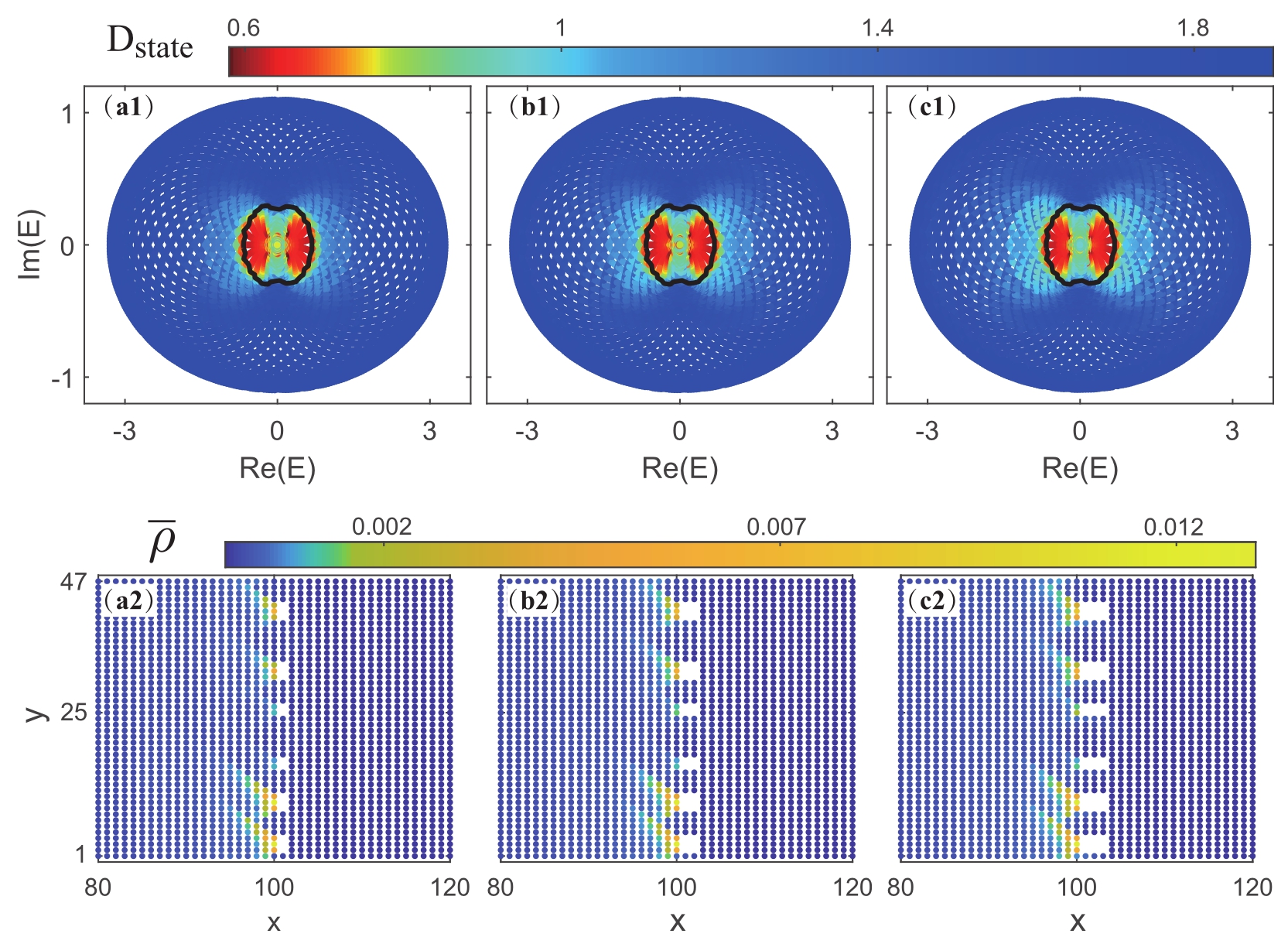}  
    \caption{Complex spectra and state distributions of the 2D system with block defects. 
    (a1) to (c1) Eigenenergies of the defective model, marked by different colors according to $D_{\rm state}$, the effective dimension of eigenstates. The parameters are $t_R=1.2$, $t_L=0.5$, $j_R=0.6$, $j_L=1$, $N_x=200$ and $N_y=47$.
    Defects with a total defect length $L_d=16$ along $y$ direction (at $y=2,3,4,9,10,11,16,17,25,26,31,32,33,41,42,43$)
    are generated by removing lattice sites from $x_d=100$ to $x_d+L_d^x$, with $L_d^x=1,2,3$, respectively.
    The black loop separates the regions with $|\overline{W}_x|$  larger and smaller than $L_0/N_y$ (with $L_0=N_y-L_d$). 
    (a2) to (c2) average distribution of all eigenstates, $\overline{\rho}=\sum_n|\psi_{x,y}^n|^2/N_y$, corresponding to (a1) to (c1), respectively.
   }
    \label{fig:block}
\end{figure}

\subsubsection{B. Skin defect states at block defects}
Next we consider defects with increased widths along $x$ direction in a 2D non-Hermitian lattice, by removing $L_d^x$ lattice sites along $x$ direction for some randomly chosen $y$ positions.
This scenario is similar to the 2D example  in Fig. 1 in the main text, but with 2D block defects aligned along $y$.
Numerical results 
with lattice sites from $x_d+1$ to $x_d+L_d^x$ being removed
for $L_d^x=1,2,3$ are demonstrated in Fig. \ref{fig:block}.
It is seen that the topological characterization of skin defect states remain valid also for such block defects; namely, they appear only at eigenenergies satisfying $|\overline{W}_x|>L_0/N_y$.


\end{widetext}
\bibliography{references}

\end{document}